\documentclass[aps, prd, amsmath, floats, floatfix, twocolumn,superscriptaddress, nofootinbib, showpacs,reprint]{revtex4}

\usepackage{hyperref}
\usepackage{amssymb,amsmath}
\usepackage{bm}
\usepackage{graphicx}
\usepackage{verbatim}
\usepackage[usenames]{color}

\newcommand{\appropto}{\mathrel{\vcenter{
  \offinterlineskip\halign{\hfil$##$\cr
    \propto\cr\noalign{\kern2pt}\sim\cr\noalign{\kern-2pt}}}}}

\newcommand{\Alfven}{Alfv\`en}

\begin{document}

\title{Interaction of misaligned magnetospheres in the coalescence of binary neutron stars}

\author{Marcelo Ponce}
\affiliation{Department of Physics, University of Guelph, Guelph, Ontario N1G 2W1, Canada}
\author{Carlos Palenzuela}
\affiliation{Canadian Institute for Theoretical Astrophysics, Toronto, Ontario M5S 3H8, Canada}
\author{Luis Lehner}
\affiliation{Perimeter Institute for Theoretical Physics,Waterloo, Ontario N2L 2Y5, Canada}
\author{Steven L. Liebling}
\affiliation{Department of Physics, Long Island University, New York 11548, USA}

\begin{abstract}
We study the dependence of the electromagnetic luminosity---produced by
interactions of force-free magnetospheres---on dipole inclinations in binary neutron
star systems.
We show that this interaction extracts kinetic energy from the system and
powers a Poynting flux with a strong dependence on the dipole orientations.
This dependence can be linked to the reconnection and redistribution 
of magnetic field                                            as the stars interact.
Although  the details of the Poynting luminosity are very much dependent on the orientation,
all the cases considered here nevertheless radiate a large Poynting flux.
This robust emission suggests
that the pre-merger stage of binary neutron star systems can yield interesting
electromagnetic counterparts to gravitational wave events.
\end{abstract}

\date{\today \hspace{0.2truecm}}


\maketitle

%%%%%%%%%%%%%%%%%%%%%%%%%%%%%%%%%%%%%%%%%%%%%%%%%%%%%%%%%%%
%%%%%%%%%%%%%%%%%%%%%%%%%%%%%%%%%%%%%%%%%%%%%%%%%%%%%%%%%%%
\section{Introduction}
%%%%%%%%%%%%%%%%%%%%%%%%%%%%%%%%%%%%%%%%%%%%%%%%%%%%%%%%%%%
%%%%%%%%%%%%%%%%%%%%%%%%%%%%%%%%%%%%%%%%%%%%%%%%%%%%%%%%%%%
Binary neutron stars are one of the primary sources of 
detectable gravitational radiation expected in the next generation
of gravitational wave interferometers LIGO/VIRGO/KAGRA~\cite{Abbott:2007kv,2011CQGra..28k4002A,Somiya:2011np}. Moreover,
these systems are also among the strongest candidates for observable neutrino production and
energetic electromagnetic output in stellar mass systems, and thus they represent
exciting possibilities for upcoming multimessenger astronomy (e.g.~\cite{Andersson:2013mrx,Kelley:2012tc,Metzger:2011bv}).

Through the combined efforts of the numerical relativity and astrophysics communities, considerable
knowledge has been gained about the expected characteristics of the gravitational
waves produced in binary neutron star mergers. Bearing in mind that a thorough 
exploration of the physical parameter space is still not at hand, several issues have 
been largely addressed.
For instance, 
much work has studied 
the dependence of gravitational waves on total mass, mass ratio, and,  to some extent, the
equation of state (see e.g.~\cite{Shibata:2005ss,Shibata:2003ga,2008PhRvD..78h4033B,Read:2009yp}). 
Recently, simulations within
full general relativity have also begun
providing important clues into the dependence on neutron star 
magnetization~\cite{2008PhRvL.100s1101A,Liu:2008xy,2011PhRvD..83d4014G,2013PhRvL.111f1105P,Lehner:2011aa,2013PhRvD..88d3011P} and 
cooling~\cite{Sekiguchi:2011zd,Sekiguchi:2011mc,2014arXiv1403.3680N}.

In addition to the dynamics and gravitational wave production,
the exploration of possible electromagnetic counterparts that complement gravitational wave observations
is particularly intriguing.
Recently we demonstrated that the interaction between the magnetospheres of each neutron star in
a binary can radiate considerable electromagnetic energy~\cite{2013PhRvL.111f1105P,2013PhRvD..88d3011P}. 
This work used a novel resistive Magneto-HydroDynamics (MHD) approach to describe the magnetic field within the disparate
regimes inside and outside each star.
Within the star, the approach adopted the ideal MHD limit while the force-free approximation
described the magnetospheres. This work started with each star having an
initial dipole field either aligned or
anti-aligned with the orbital angular momentum and found
a powerful Poynting luminosity produced as the stars orbited.
Such a large luminosity provides the tantalizing possibility
of powering electromagnetic counterparts to gravitational waves from the system.
The Poynting flux for equally magnetized stars with either aligned or anti-aligned
dipole moments was found to be collimated along the polar region. In contrast, when
one star was unmagnetized or barely magnetized with respect to the other,
the Poynting flux was mainly directed around the equatorial region.
Naturally, generic systems are not expected to have such a preferred alignment of
dipole magnetic moments, and thus the study of different scenarios is important.
We address this question here by considering
misaligned dipoles in binary neutron star mergers. 

According to the standard formation channel, a binary neutron star system is
formed from a primordial binary through a sequence of complex processes
\cite{lrr-2012-8,Lorimer:2008se,Kalogera:2006uj,1991PhR...203....1B}.
The first of these processes is
a supernova explosion of the more massive star once it evolves off
the main sequence and through its giant phase.
The remnant of such an explosion becomes the first NS of the binary, generally
more massive than the second NS and with a potential recoil. Subsequently,
the secondary star then evolves off the main sequence and,
ultimately, explodes as a supernova becoming the second NS.

The discovery of strongly relativistic, binary pulsars provides strong support
for the formation channel described above.
Any kicks provided by the supernovae are important not only in determining whether the binary
survives, but also in determining the properties of the resulting binary.
In particular, kicks can tilt the orbital plane of the binary and misalign the individual spins of the NSs \cite{Kalogera:1999tq}.
Furthermore, it has been empirically argued~\cite{2008MNRAS.387.1755W,2010MNRAS.402.1317Y}
that the angle between the magnetic and spin axes of NSs may not be random
but instead correlated to each other.
The orientations of the magnetic moments in generic binaries are therefore plausibly arbitrary.

In this work, we consider different orientations of the stellar magnetic dipoles
and compute the resulting Poynting flux characteristics as the stars 
coalesce\footnote{Tilted magnetic fields have also been considered in
black hole-neutron star binaries to asses their role in inducing toroidal magnetic configurations
in the resulting accretion disk~\cite{Etienne:2012te}.}. In addition to
a strong flux of electromagnetic energy produced due to magnetosphere interactions,
our results indicate that one can bracket the expected luminosities from general configurations.
Our simulations focus primarily on the last orbits before the merger in which the dynamics are most
rapid and violent and  simplified analytic models are not applicable.
Additional insight comes from modeling the behavior of a magnetized star moving within
the field of a distant star, modeled here as a constant external field.
While such an approach is only valid when the stars
are well separated, it serves to generalize well-known
results with the unipolar inductor.

This work is organized as follows, in Sec.~\ref{section:model} we present
theoretical arguments and estimates for the possible electromagnetic luminosity induced
by the system.
Sec.~\ref{section:methods} summarizes briefly the evolution equations describing the magnetized
neutron stars, as well as our numerical setup.
Our results are presented in Sec.~\ref{section:results},
followed by conclusions in Sec.~\ref{section:discussion}.

%%%%%%%%%%%%%%%%%%%%%%%%%%%%%%%%%%%%%%%%%%%%%%%%%%%%%%%%%%%
%%%%%%%%%%%%%%%%%%%%%%%%%%%%%%%%%%%%%%%%%%%%%%%%%%%%%%%%%%%
\section{Theoretical estimates}
\label{section:model}
%%%%%%%%%%%%%%%%%%%%%%%%%%%%%%%%%%%%%%%%%%%%%%%%%%%%%%%%%%%
%%%%%%%%%%%%%%%%%%%%%%%%%%%%%%%%%%%%%%%%%%%%%%%%%%%%%%%%%%%

To estimate the power of the systems studied here,
we begin by considering the process of {\em unipolar
induction}~\cite{1969ApJ...156...59G}. Within the simple model discussed
in Ref.~\cite{1965PhRvL..14..171D}, 
the electromagnetic power radiated by an unmagnetized, perfectly
conducting neutron star moving 
through a magnetically dominated medium (i.e., such that
the \Alfven\, speed approaches the speed of light) can be estimated
by
\begin{equation}
  \mathcal{L}_{\rm Alfven} \approx \frac{v^2}{2 c} B_z^2 R_c^2,
\label{eq:alfven_lum}
\end{equation}
where $v$ is the relative velocity between the neutron star and
the external magnetic field of strength $B_z$, and $R_c$ is the
radius of the neutron star. 

We apply this estimate to a binary, in which we identify the
moving star as the unmagnetized companion. The primary star in
this system is then the source of the magnetic field external to
the companion. As such, we assume a dipole field for the
primary so that $B_z=B_* (R_*/a)^3$ where $B_*$ is the magnetic
field strength at the pole of the primary, $R_*$ its radius, and
$a$ the binary separation.  Substitution for $B_z$ into
Eq.(\ref{eq:alfven_lum}) yields~\cite{2001MNRAS.322..695H}
\begin{equation}
  \mathcal{L}_{\rm Alfven} \approx  \left(\frac{v}{c}\right)^2
      \frac{B_*^2 R_*^6 R_c^2 c}{2 a^6} ~~,
\label{eq:alfven_lum_binary}
\end{equation}
which is consistent with
 the dissipation rate estimated 
in Ref.~\cite{Lai:2012qe} by using a circuit model for the magnetic
interactions of such a binary.

Although this estimate applies only to binaries with an unmagnetized
companion, neutron stars in binaries are unlikely to have a vanishing
magnetic moment.
It is however  possible to generalize the previous luminosity estimate
to systems with a relatively weaker
magnetized companion with respect to the primary's magnetization as follows. 
Consider the companion having a (dipolar) magnetic field. In the electrovacuum case, the strength 
of this field will dominate over the external one (induced by the primary) roughly
when the following condition (centered at the location of the companion) is satisfied
\begin{equation}
   B_c \left ( \frac{R_c}{r} \right )^3 \sqrt{\frac{1 + 3 \sin^2(\pi/2-\theta)}{4}} > B_z 
\label{eq:dominance_region}
\end{equation}
with $B_c$ the magnetic strength at the pole of the companion and
$\theta$ the angle between the stellar magnetic moment
and the radial vector to a given point. Also, for simplicity $B_z$ is the strength of a constant field,
representing the effect of a widely separated primary star.
Neglecting the angular dependence,
we call the surface
around the companion where the internal magnetic field balances the external field
a {\em screening sphere} for the following calculation. 
The radius $R_m$ of this screening
sphere is just
\begin{equation}
         R_m \approx R_c s_z^{1/3}~~~,~~ s_z \equiv B_c/B_z.
\label{eq:effective_radius}
\end{equation}
This is applicable as long as $R_m \gg R_c$ or $s_z \gg 1$, but in
the case of an unmagnetized companion one must set 
$R_m = R_c$ in order to recover the \Alfven\, radiation
power of Eq.(\ref{eq:alfven_lum}). 
Thus, to account for both limits, we slightly modify Eq.(\ref{eq:effective_radius}) to
recover the appropriate values for both $s_z \gg 1$ and
$s_z \ll 1$
\begin{equation}
R_m \equiv  R_c \max\left(1, s_z^{1/3}\right).
\label{eq:effective_radiusMOD}
\end{equation}
We now substitute $R_m$ for the radius of the companion $R_c$ in
Eq.(\ref{eq:alfven_lum}) to get  an estimate of the \Alfven\,
radiation power of a magnetized star moving through an external field
\begin{equation}
  \mathcal{L}_{\rm Alfven} \approx
  \frac{v^2}{2 c} B_z^2 R_c^2 \max\left(1, s_z^{2/3}\right).
\label{eq:alfven_lum2}
\end{equation}

We stress that this estimate relies on several approximations.
First, we have neglected the angular dependence in
Eq.~(\ref{eq:dominance_region}), which might modify the screening radius
by at most a factor $2$ (i.e., when the magnetic moment of the companion
is perpendicular to the external magnetic field). Second,
Eq.~(\ref{eq:effective_radiusMOD}) is only valid in the asymptotic limits
$s_z \gg 1$ and $s_z \ll 1$, and so deviations in the luminosity are naturally expected
when $s_z \approx 1$. Finally, the luminosity in Eq.~(\ref{eq:alfven_lum2})
results from rather simple arguments which, for example,  do not consider
reconnection or deformation of field lines. 
We expect that these corrections will not modify the
dependence $s_z^{2/3}$ for large $s_z$, but,
as we show via simulations later, the intensity of the resulting luminosity
can be affected by their impact on the proportionality factor (see e.g.
Fig~\ref{fig:L_betav}).

Returning to a binary system, again neglecting the angular dependence,
we can similarly equate the fields of the primary and companion
\begin{equation}
   B_* R_*^3/\left(a - R_m\right)^3 = B_c \left(R_c/R_m\right)^3.
\label{eq:effective_radius2}
\end{equation}
Assuming that the two stars have similar radii $R_* \approx R_c$ and that
they are far apart $a \gg R_m$, Eq.(\ref{eq:effective_radius2}) leads to
$R_m \approx a (B_c/B_*)^{1/3}$. 
Again placing a lower limit on the effective radius, we have
$R_m = \max [R_c, a (B_c/B_*)^{1/3}]$. We generalize Eq.(\ref{eq:alfven_lum_binary})
obtaining the \Alfven\, radiation power for a magnetized binary by 
substituting $R_m$ for $R_c$.
For small separations, $R_m \approx R_c$ and this generalized power hardly
differs from Eq.(\ref{eq:alfven_lum_binary}). However, for large separations 
the magnetic field of the primary will be small near the companion, and in
that limit the luminosity will be
\begin{equation}
  \mathcal{L}_{\rm Alfven} \approx  \left(\frac{v}{c}\right)^2
      \frac{B_*^2 R_*^6 s_*^{2/3} c}{2 a^4} ~~, s_* \equiv B_c/B_* ~~,
\label{eq:alfven_lum_binary2}
\end{equation}
which enhances the original \Alfven\, power by a factor $(a^2/R_c^2) s_*^{2/3}$.

Let us stress again that this result is only valid for 
companions with weaker magnetizations (with respect to the primary) and for binaries in a quasi-adiabatic stage.
Assuming Keplerian orbits $v^2 \propto a^{-1}$ for simplicity, we can see from 
Eq.(\ref{eq:alfven_lum_binary}) that when the companion is unmagnetized the luminosity scales as $a^{-7}$.
On the other hand, when the two stars are magnetized, the radiated electromagnetic luminosity  behaves
differently because of the interaction between each star with the field of the other. 
This interaction depends strongly on the relative orientation of the dipole fields of the stars, 
as we discuss later. For weakly magnetized companions, it follows from Eq.(\ref{eq:alfven_lum_binary2}) that 
the luminosity has a softer dependence $a^{-5}$. Finally, for comparable magnetizations and with
dipole moments aligned or antialigned, we found in previous work~\cite{2013PhRvL.111f1105P,2013PhRvD..88d3011P}
that in the last orbits $\mathcal{L} \appropto a^{-3}-a^{-1.5}$. These three estimates
demonstrate  a clear influence
by the stellar magnetization on the binary luminosity.  Notice as well that as the orbit tightens and the
stellar accelerations increase, an additional contribution (a ``Larmor-type'' term)  
becomes non-negligible~\cite{Brennan:2013ppa} and may contribute to soften
the luminosity dependence on $a$.

%%%%%%%%%%%%%%%%%%%%%%%%%%%%%%%%%%%%%%%%%%%%%%%%%%%%%%%%%%%
%%%%%%%%%%%%%%%%%%%%%%%%%%%%%%%%%%%%%%%%%%%%%%%%%%%%%%%%%%%
\section{Numerical approach}
\label{section:methods}
%%%%%%%%%%%%%%%%%%%%%%%%%%%%%%%%%%%%%%%%%%%%%%%%%%%%%%%%%%%
%%%%%%%%%%%%%%%%%%%%%%%%%%%%%%%%%%%%%%%%%%%%%%%%%%%%%%%%%%%

Our aim is to study the behavior of magnetically dominated
plasma surrounding single or binary magnetized neutron stars.
The complexity of these problems demands numerical simulations
that incorporate  general relativity and relativistic, resistive
magnetohydrodynamics~\cite{2013MNRAS.431.1853P} (see~\cite{2013PhRvD..88j4031P}
for a related approach not relying on resistive MHD).
We have developed such a framework and applied it recently
to study the collapse of neutron stars~\cite{2013MNRAS.431.1853P} and
magnetospheric interactions of magnetized binary neutron stars
with aligned/anti-aligned magnetic
moments~\cite{2013PhRvL.111f1105P,2013PhRvD..88d3011P}.

Our approach solves the Einstein-Maxwell-hydrodynamic
equations to model strongly gravitating compact stars and the
effects of a global electromagnetic field. Inside the star, the magnetic
field is modeled within the ideal MHD limit, transitioning to the
force-free limit outside the stars. This transition is achieved,
in our resistive MHD framework, by a prescription for the
(anisotropic) conductivity tensor that depends on the fluid density
(see~\cite{2013MNRAS.431.1853P} for details).
We adopt the BSSN formulation~\cite{Baumgarte:1998te,Shibata:1995we} 
of the Einstein equations as described in~\cite{Neilsen:2010ax}.

We use finite difference and finite volume techniques on a regular,
Cartesian grid to discretize
the system~\cite{SBP2,SBP3}. The geometric fields are discretized
with a fourth order accurate scheme satisfying the summation by
parts rule, while High Resolution Shock Capturing methods based on the HLLE
flux formulae with PPM reconstruction are used to discretize the fluid and
the electromagnetic variables~\cite{Anderson:2006ay,Anderson:2007kz}.
The time evolution of the resulting equations must deal with the presence
of stiff terms arising from the resistive MHD scheme in the evolution of
the electric field in Maxwell's equations. Such terms are efficiently handled
with a third order accurate IMEX (implicit-explicit) Runge-Kutta scheme,
as described in~\cite{Palenzuela:2008sf,2013PhRvD..88d4020D,2013MNRAS.431.1853P}. 

To ensure sufficient resolution is available in an efficient manner, we employ
adaptive mesh refinement~(AMR) via the HAD computational infrastructure that
provides distributed, Berger-Oliger style AMR~\cite{had_webpage,Liebling} with
full sub-cycling in time, together with an improved treatment of artificial
boundaries~\cite{Lehner:2005vc}. In this work we only use fixed mesh refinement
(FMR), with refinement boxes covering the regions surrounding the stars.
The grid resolutions adopted here are the same as those used previously
to obtain convergent evolutions~\cite{2013PhRvL.111f1105P,2013PhRvD..88d3011P}. We
confirm this convergence here by varying the resolution by $20\%$ for one of the
most extreme binary configurations (i.e., the $P/P$ case defined in 
Section~\ref{section:results_binary})
and finding agreement in the obtained luminosities to within $\approx 2\%$.

%%%%%%%%%%%%%%%%%%%%%%%%%%%%%%%%%%%%%%%%%%%%%%%%%%%%%%%%%%%
%%%%%%%%%%%%%%%%%%%%%%%%%%%%%%%%%%%%%%%%%%%%%%%%%%%%%%%%%%%
\section{Results}
\label{section:results}
%%%%%%%%%%%%%%%%%%%%%%%%%%%%%%%%%%%%%%%%%%%%%%%%%%%%%%%%%%%
%%%%%%%%%%%%%%%%%%%%%%%%%%%%%%%%%%%%%%%%%%%%%%%%%%%%%%%%%%%

Here we describe the details of our numerical studies of 
magnetized neutron stars. We begin with a magnetized star boosted 
with respect to an external magnetic field. It is natural
to generalize the analysis of this
system to the case
of a binary system with a magnetized primary
and a weakly magnetized companion. We then study the
effects of varying the orientation of stellar magnetic moments
(with equal magnitude), on the resulting Poynting luminosity
in a binary system.

\subsection{Boosted Magnetized Neutron Star}
\label{section:results_boosted}
%---------------------------------------------
\begin{figure*}
\begin{center}
   \includegraphics[width=.24\textwidth]{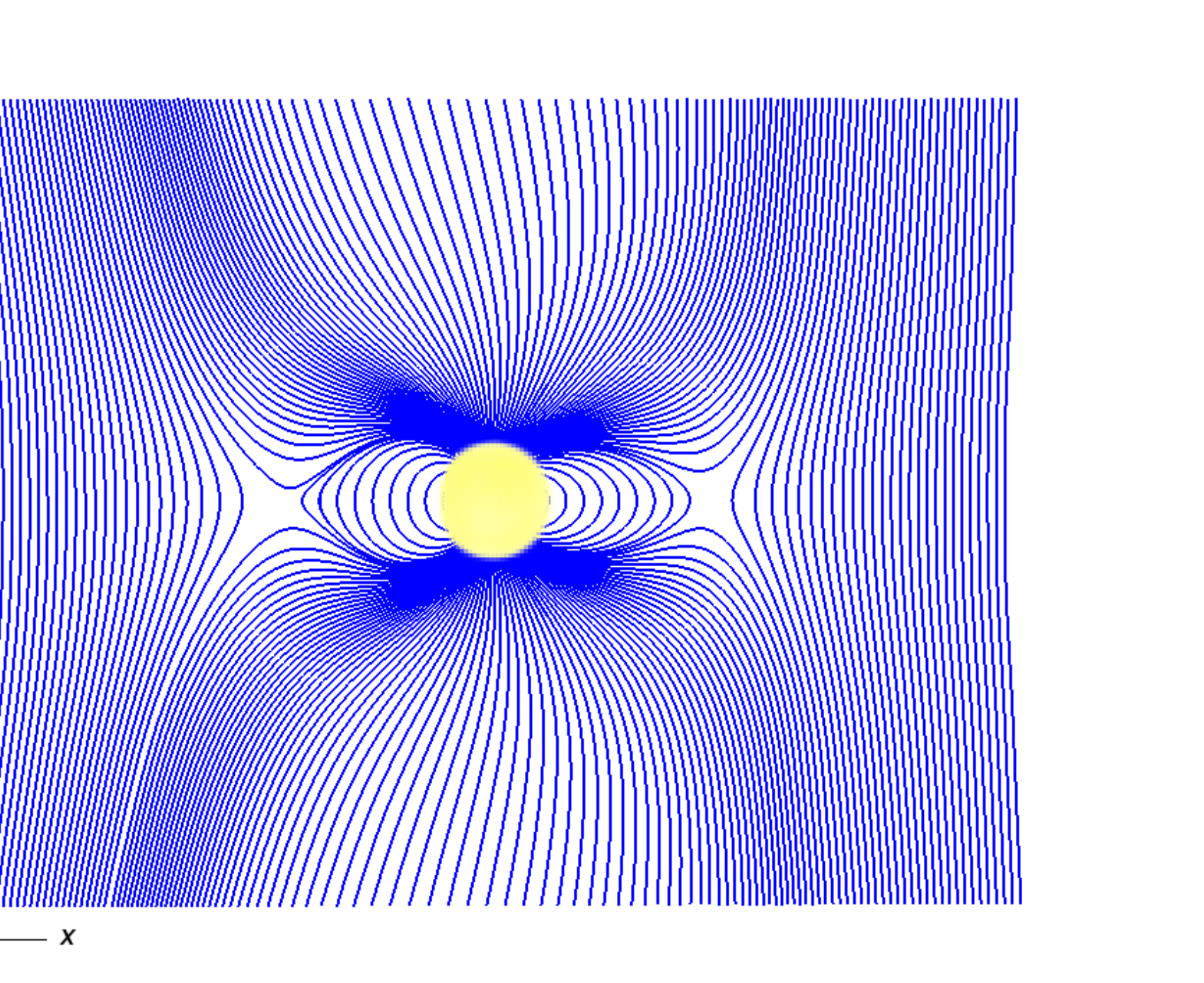}
   \includegraphics[width=.24\textwidth]{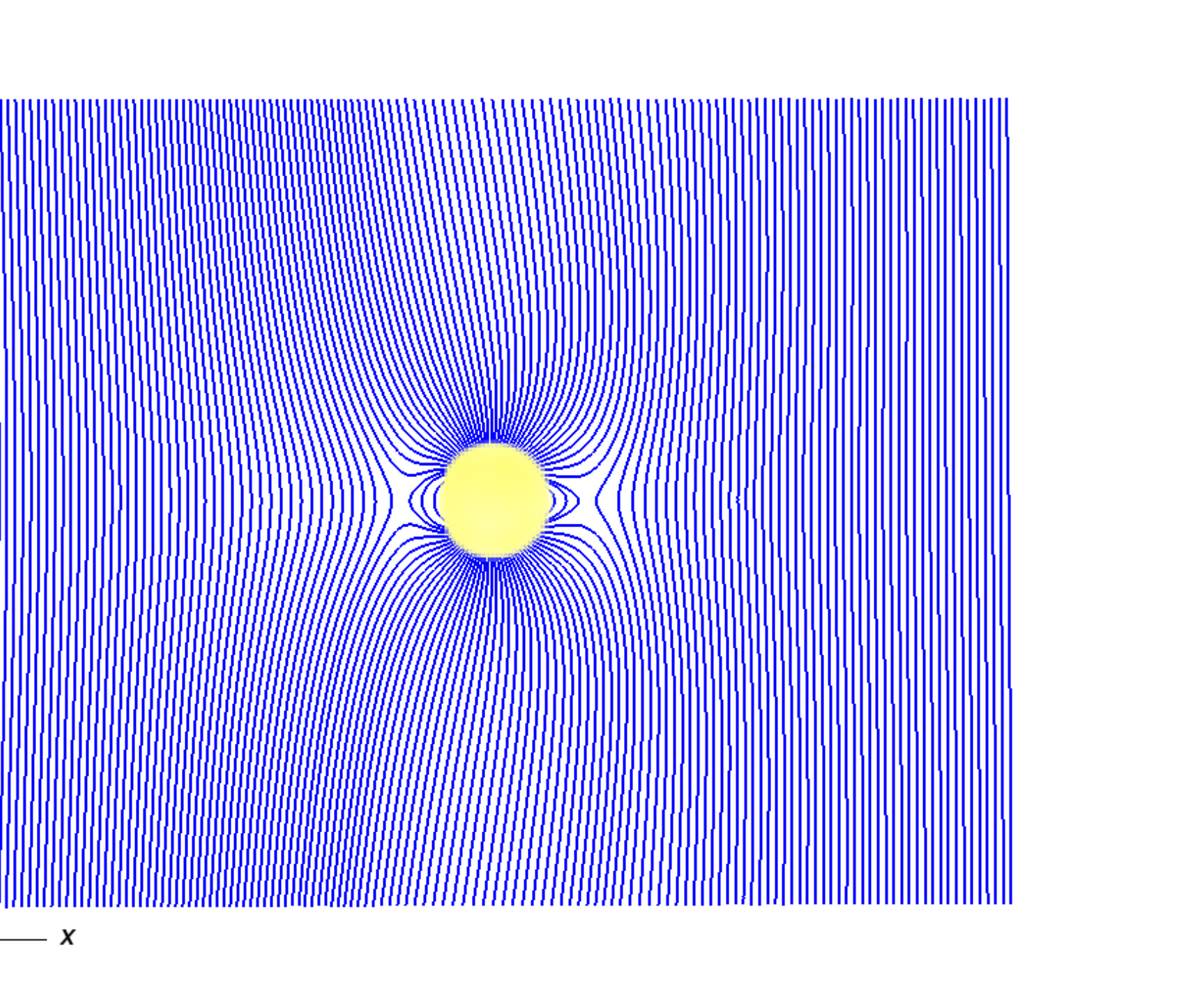}
   \includegraphics[width=.24\textwidth]{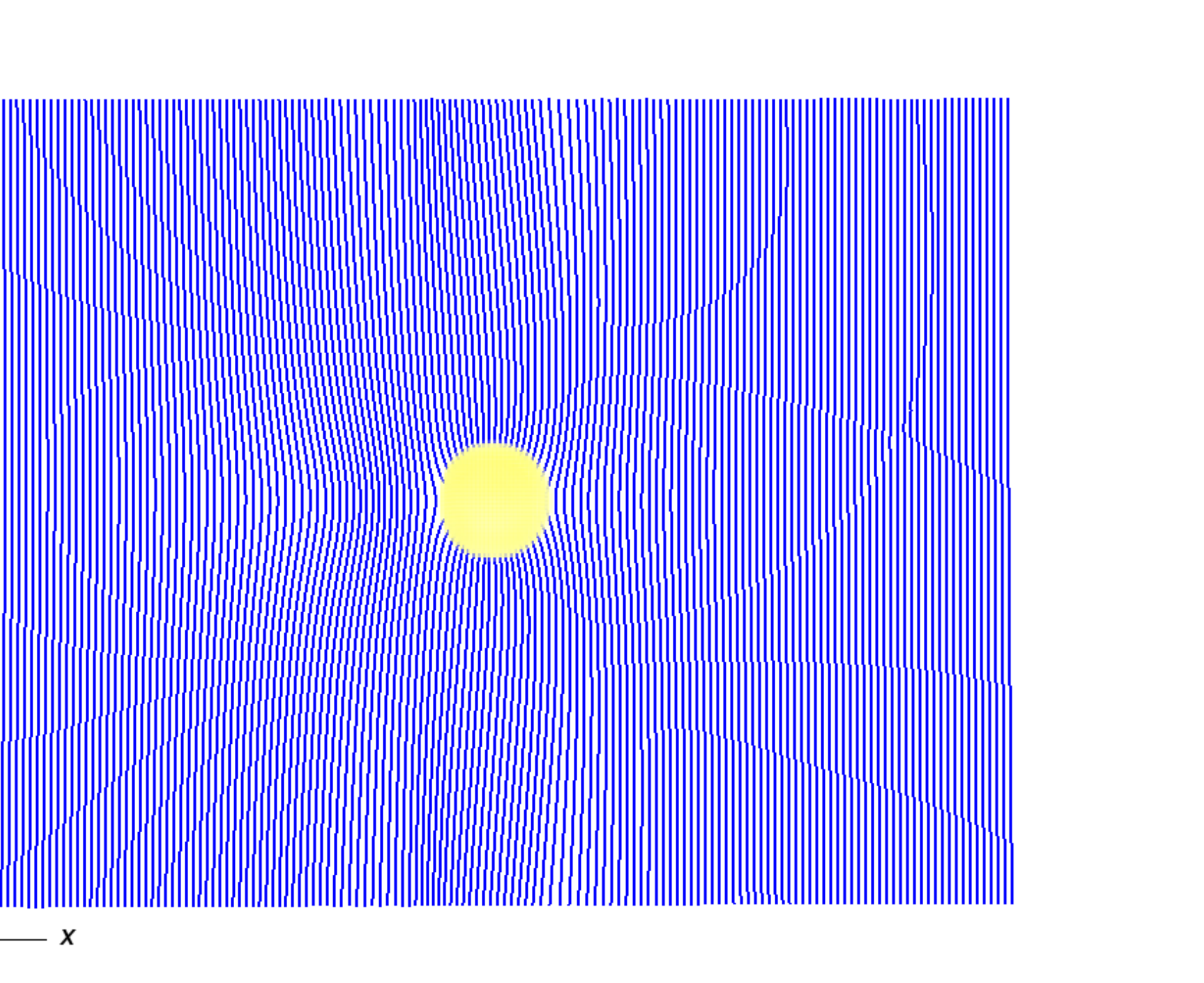}
   \includegraphics[width=.24\textwidth]{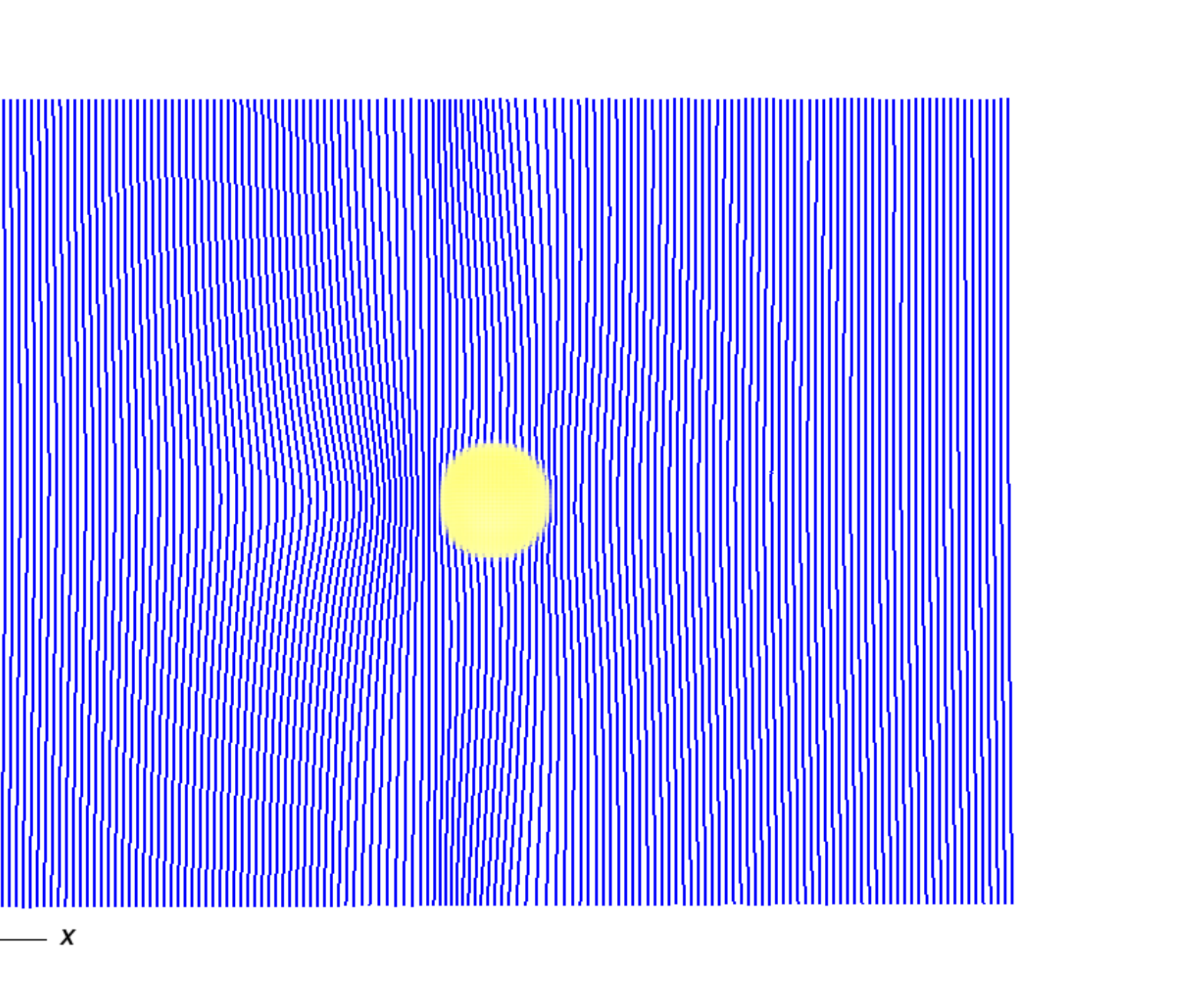}\\
   \includegraphics[width=.24\textwidth]{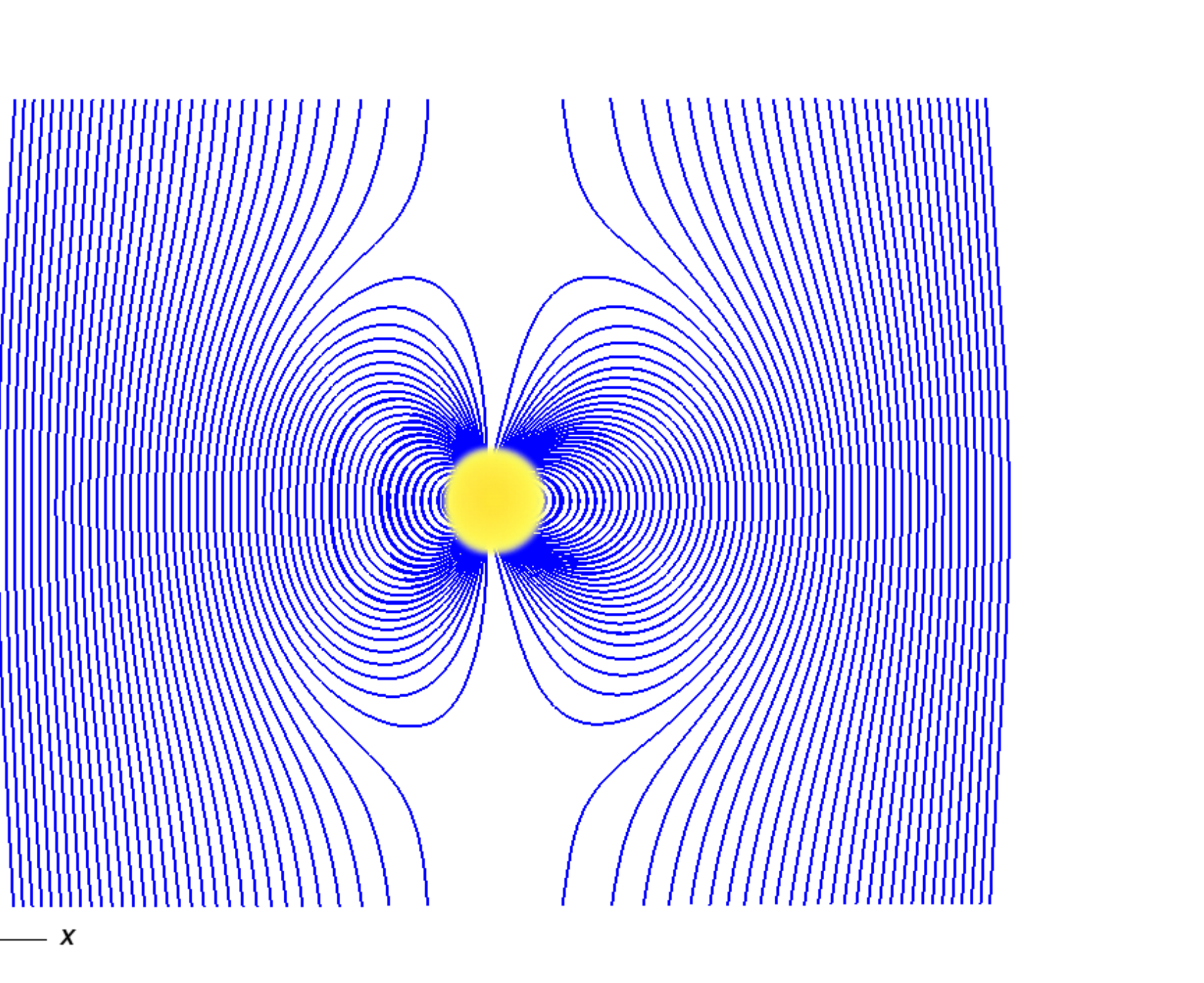}
   \includegraphics[width=.24\textwidth]{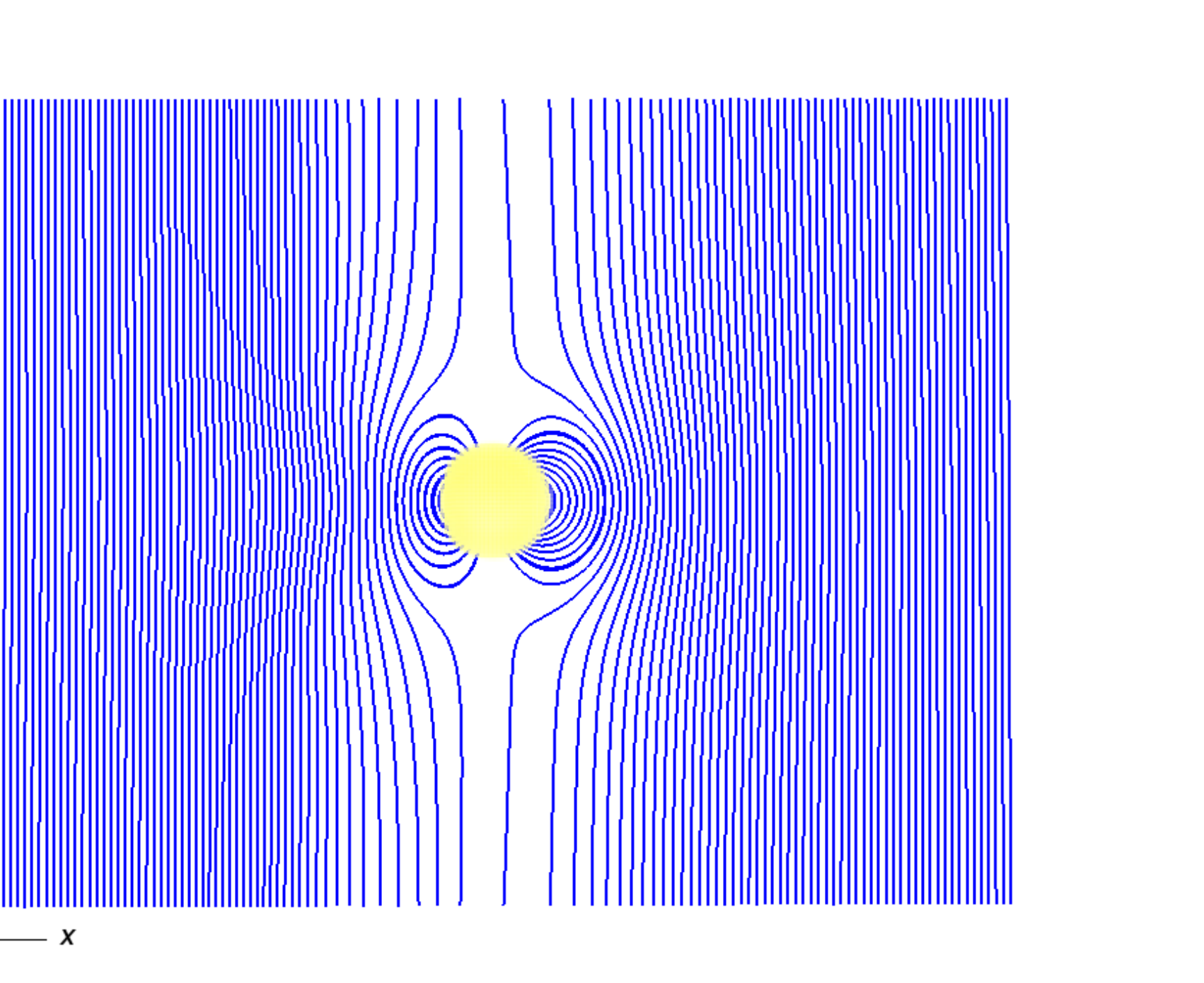}
   \includegraphics[width=.24\textwidth]{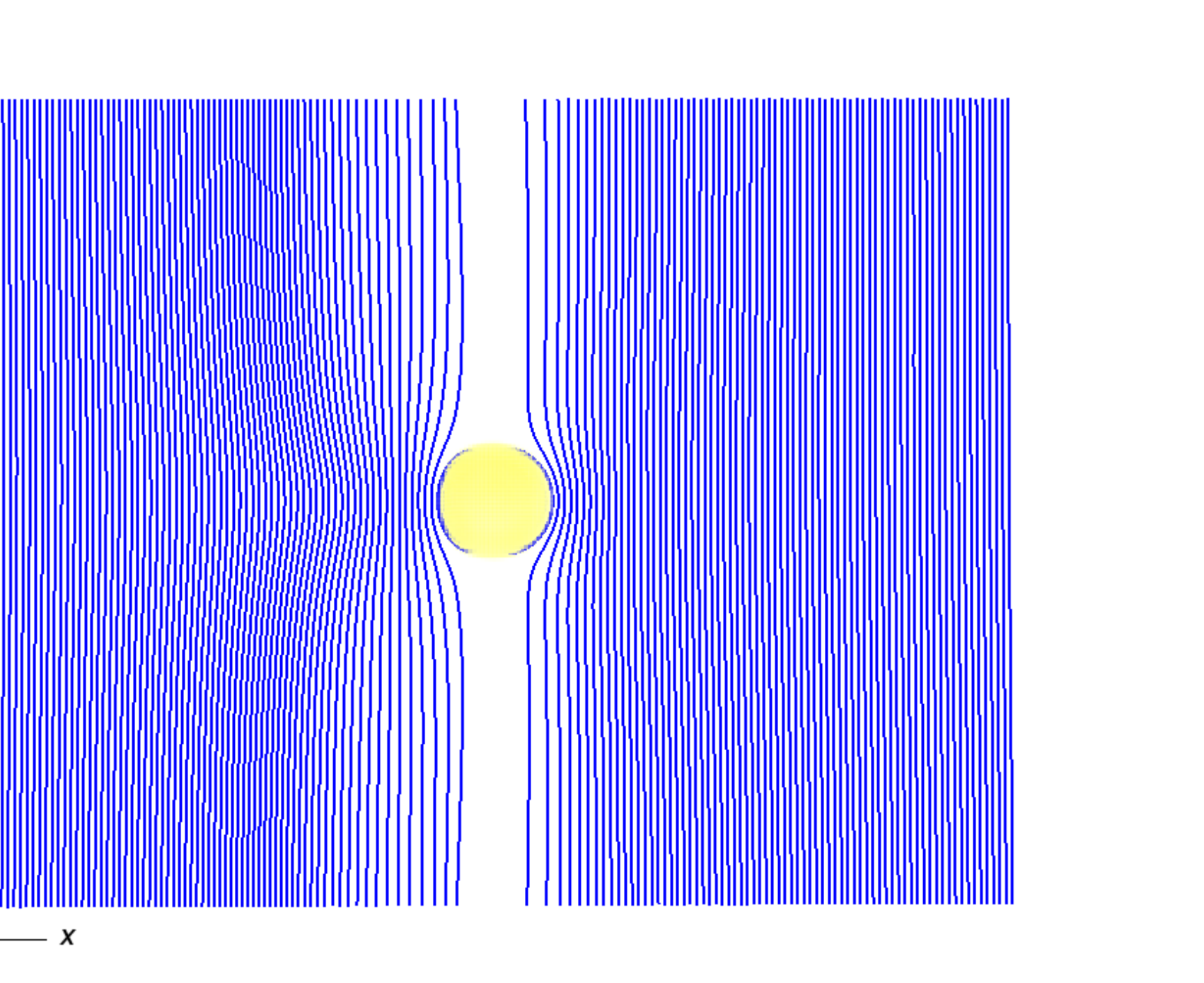}
   \includegraphics[width=.24\textwidth]{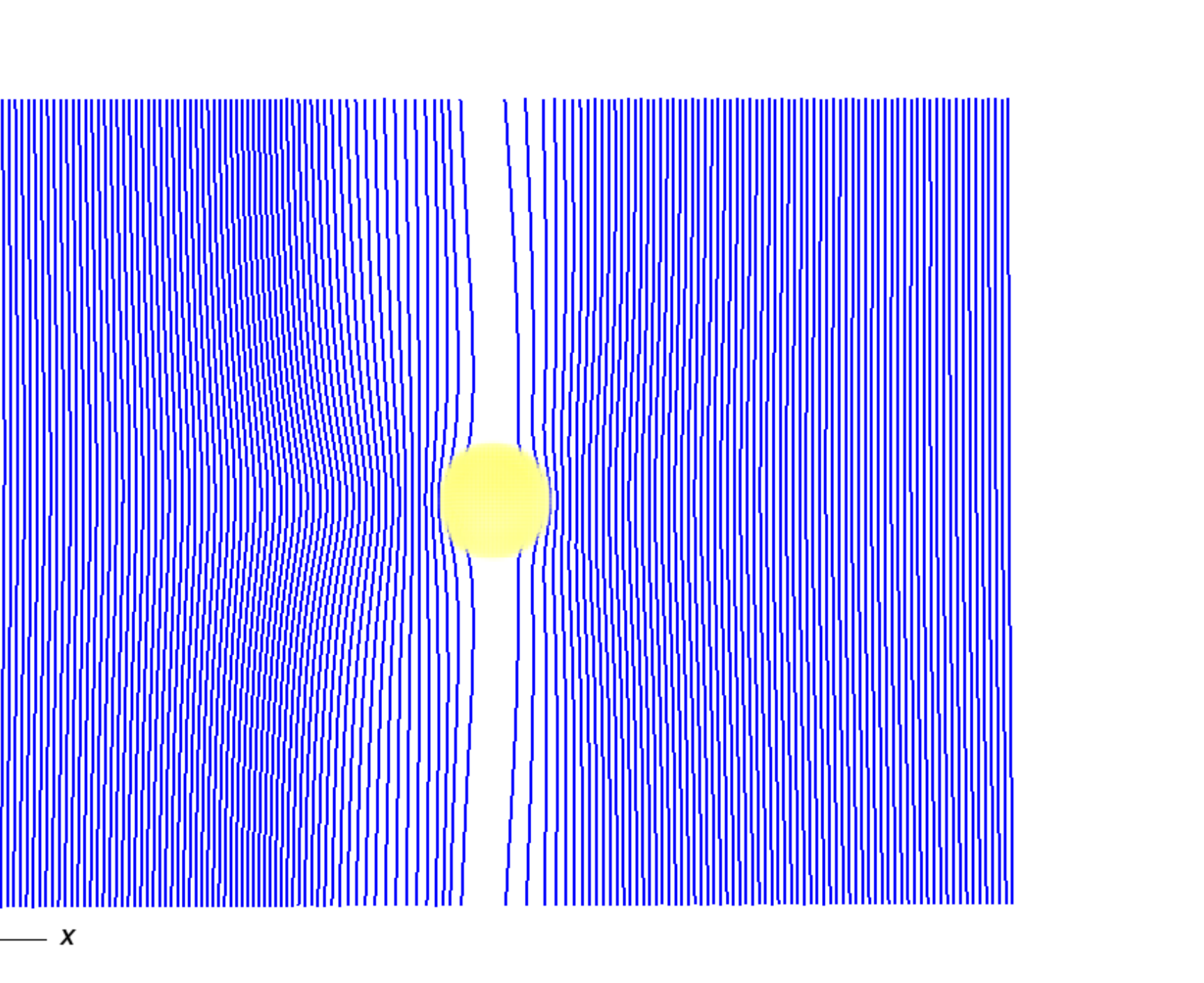}
   \caption{Magnetic field lines (blue) and stellar density (yellow) for a star boosted to the right 
     with speed $v/c=0.05$, once the system reaches a quasi-stationary state ({ 
     which depends primarily on the velocity of the star; roughly $5$ms for these cases}).
     In the {\bf top} panel, the magnetic moment is anti-aligned with an external magnetic field
     allowing for reconnection in the polar regions above and below the star.
     The initial magnetic strength ratio is given by
     $s_z \equiv B_c/B_z=\{-100,-10,-1,-0.1\}$ from left to right.
     In the {\bf bottom} panel, the stellar magnetic moment is aligned with an external
     magnetic field, leading to opposing field lines in the polar regions
     and a Poynting luminosity smaller than the anti-aligned case
     for the same magnitudes $s_z=\{100,10,1,0.1\}$ (see the top panel of Fig.~\ref{fig:L_betav}).
    } \label{fig:Bfield_beta}
\end{center}
\end{figure*}

%---------------------------------------------

%---------------------------------------------
\begin{figure}
\begin{center}
      \includegraphics[width=.75\columnwidth]{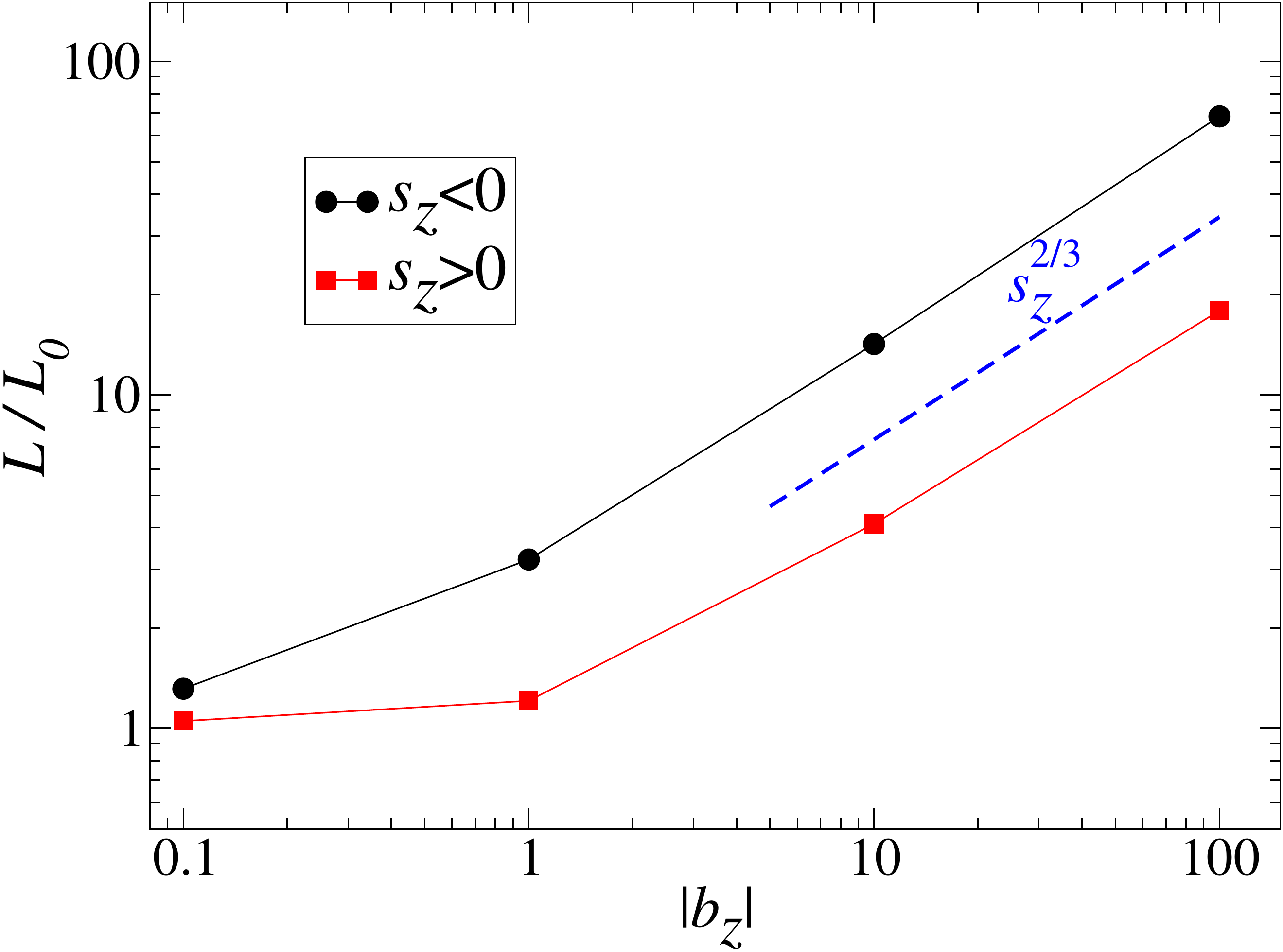}
	\\
      \includegraphics[width=.75\columnwidth]{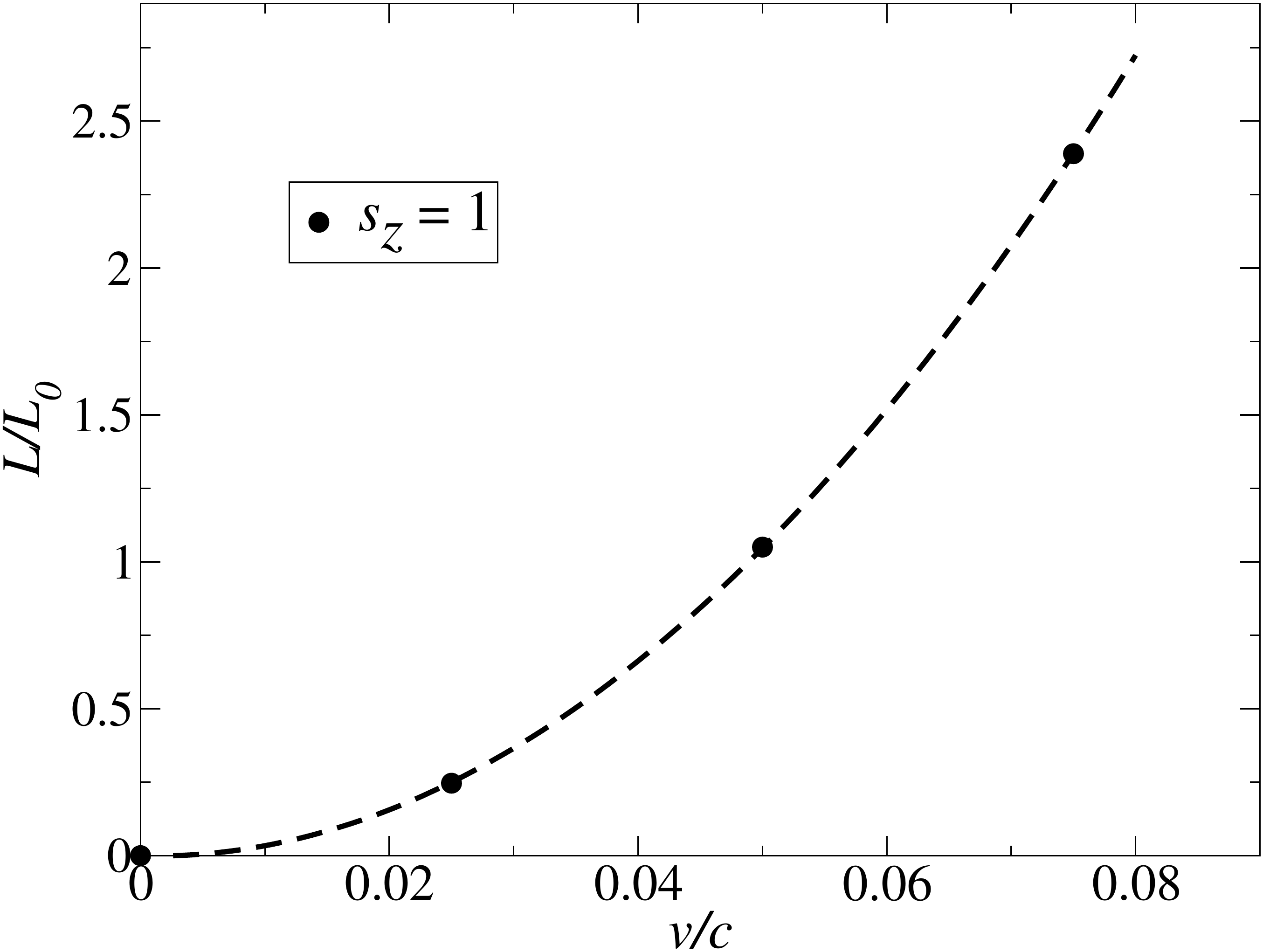}
      \caption{ \textit{Top-panel}: Luminosity as a function of $s_z\equiv B_c/B_z$ for 
       a boosted star with $v/c=0.05$ for both anti-aligned and aligned magnetic
       dipoles. Notice that while both cases have approximately the same dependence on $s_z$, the
       anti-aligned case is more luminous.
       \textit{Bottom-panel}: Luminosity as a function of the velocity of the boosted star
       with $s_z=1$ (aligned case). Also shown is a fit (dashed curve) assuming a
       quadratic dependence on velocity ($\propto v^2$).
       For concreteness we have normalized the luminosities in both plots by
        $L_0 \equiv L(s_z=0, v/c=0.05)$.
       } \label{fig:L_betav}
\end{center}
\end{figure}

%---------------------------------------------

Consider a non-rotating, magnetized neutron star 
moving with constant velocity
with respect to a uniform, external magnetic field.
This system models a weakly magnetized star moving slowly within
the field produced by a distant, primary star.
Because the emitted luminosity can be rescaled by the external
magnetic field $B_z$, a convenient way to parameterize the results
is with the dimensionless ratio $s_z \equiv B_c/B_z$, where $B_c$ is
the magnetic field strength at the pole of the boosted star. For simplicity,
we assume that the stellar magnetic moment is either aligned or
anti-aligned with 
the external, asymptotically constant magnetic field $B_z$.

We evolve  with a range of magnetizations $|s_z|=\{100,10,1,0.1\}$ and
monitor the system (luminosity, field topology, etc.)
until the star reaches a quasi-steady configuration. Examples of the relaxed
configurations are shown in
Fig.~\ref{fig:Bfield_beta} for $s_z<0$ (top)
and $s_z>0$ (bottom).
The figure makes clear that  positive and negative values of $s_z$ yield
significantly different behavior. For negative $s_z$, the polar magnetic field reconnects
with the external field, effectively connecting the star magnetically to
the asymptotic field. As a consequence of this connection a direct path for \Alfven \,
radiation is induced as the stellar motion disrupts the external field.

In contrast, for positive $s_z$, the direction of the stellar field at the polar region
is opposite to  that of the external field. This opposition creates a roughly
spherical region in which the external field is screened. 
The resulting deformation of the field lines around this region produces
\Alfven \, waves with a released power smaller than the negative case.

The total luminosity as a function of
$|s_z|$ is shown in the top panel of Fig.~\ref{fig:L_betav}. As
anticipated analytically, it displays a polynomial growth dependence
as $|s_z|^{2/3}$ for $s_z \gg 1$, and reduces to the unmagnetized
case for $s_z \ll 1$. Around $s_z \approx 1$ neither of these limits
is well satisfied and we expect departures from our simplified expression.
Interestingly, a stronger luminosity results for $s_z < 0$, suggesting that the 
continuous reconnection occurring in the stellar polar regions yields
a more powerful Poynting flux.
We also confirm that the \Alfven\, power is proportional
to $v^2$ by studying a magnetized star with fixed $s_z=1$ and different boost
velocities $v/c=\{ 0.025, 0.05, 0.075 \}$, as shown in the bottom
panel of Fig.~\ref{fig:L_betav}. 
These velocities have been chosen
bearing in mind that  when stars in a binary are separated by about $400 {\rm km}$ their velocities are roughly $v/c \sim 0.1$.
At this (or larger) separation, to leading order the star can be regarded as moving through
the field produced by the other without strong, non-linear effects playing a role in the dynamics.

%%%%%%%%%%%%%%%%%%%%%%%%%%%%%%%%%%%%%%%%%%%%%%%%%%%%%%%%%%%
\subsection{Binary neutron stars}
\label{section:results_binary}
%%%%%%%%%%%%%%%%%%%%%%%%%%%%%%%%%%%%%%%%%%%%%%%%%%%%%%%%%%%

%--------------------------------------

\begin{figure}
	\includegraphics[width=.9\columnwidth]{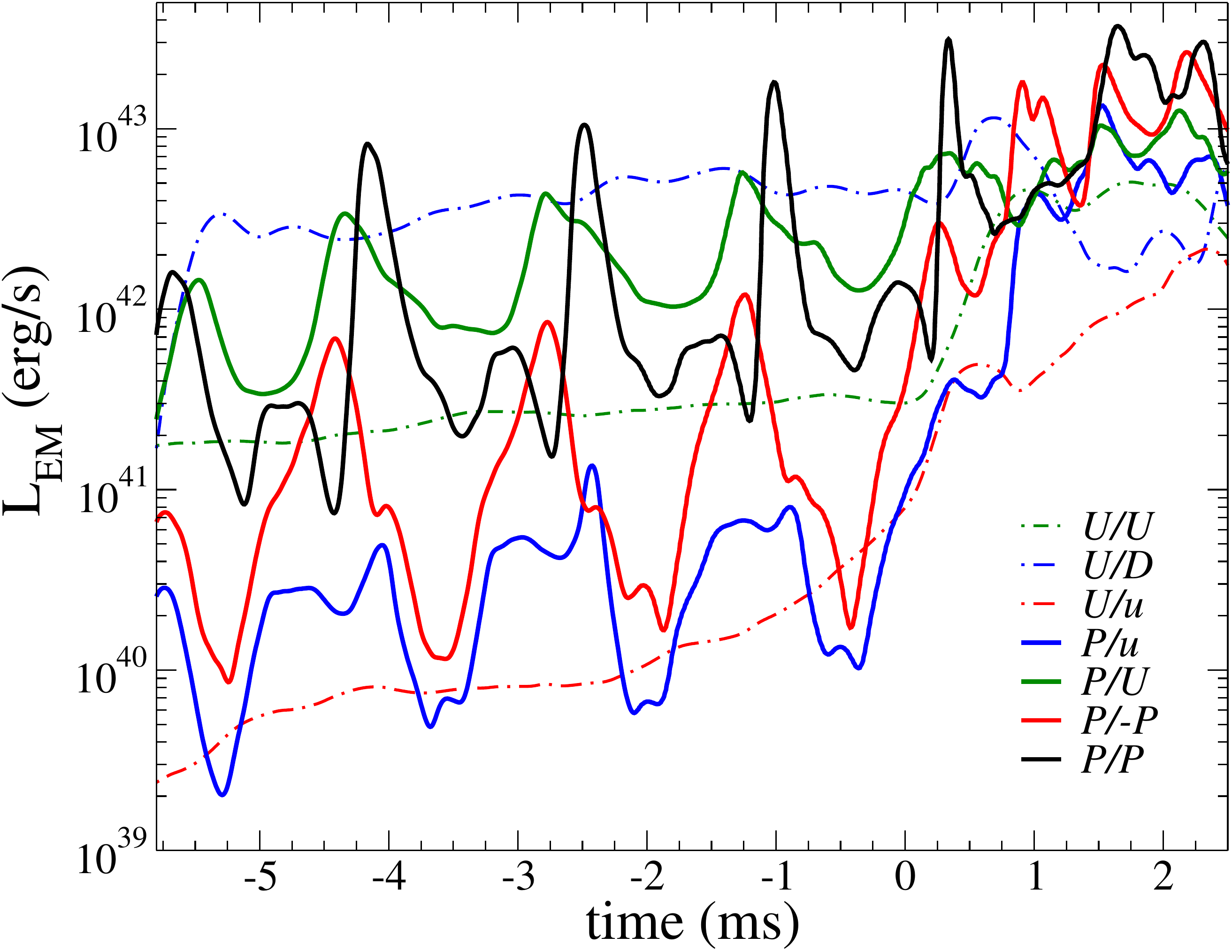}
	\caption{Luminosities for different magnetic dipole configurations in a binary
                 neutron star system, shifted in time such that $t=0$ denotes first
                 contact between the stars. The configurations studied here are displayed
                 with solid lines, whereas those studied previously in~\cite{2013PhRvL.111f1105P,2013PhRvD..88d3011P}
                 are also included (dot-dashed lines) for reference.
                 Note that the $U/u$ and $U/D$ configurations roughly bracket the values obtained
                 in the misaligned cases.
	}
        \label{fig:lums}
\end{figure}

\begin{figure*}[h!]
        \includegraphics[width=.24\textwidth]{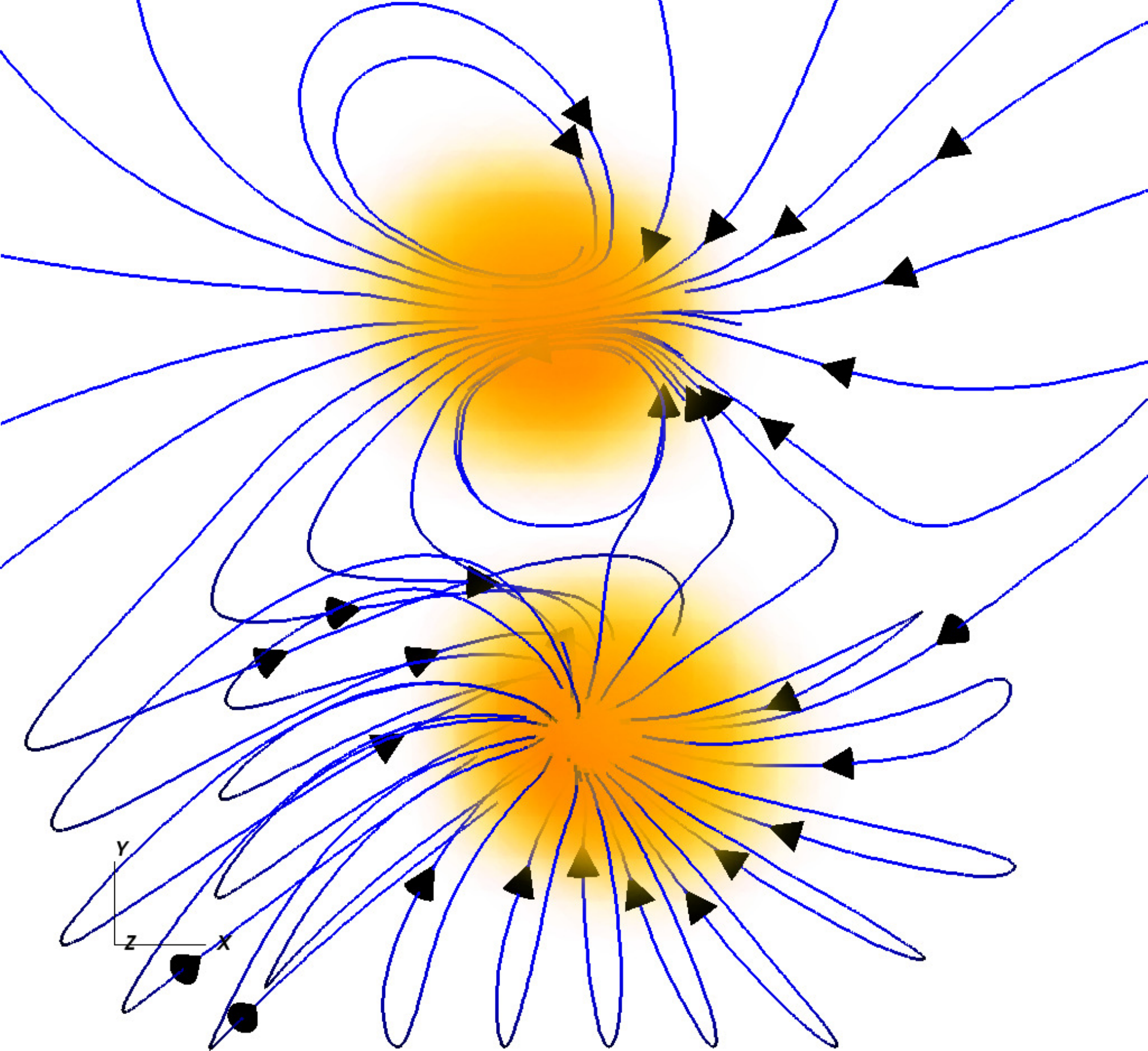}
        \includegraphics[width=.24\textwidth]{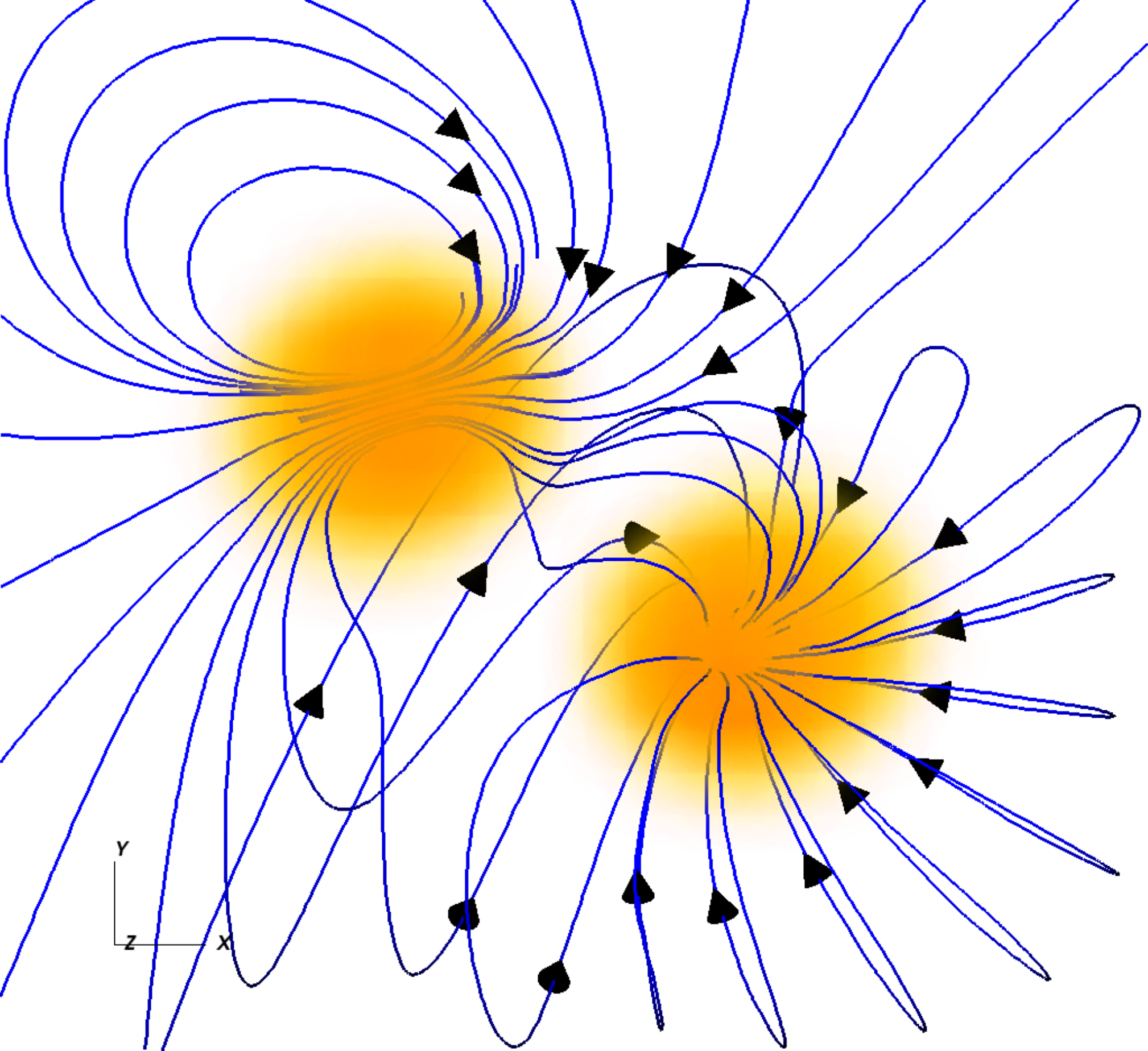}
        \includegraphics[width=.24\textwidth]{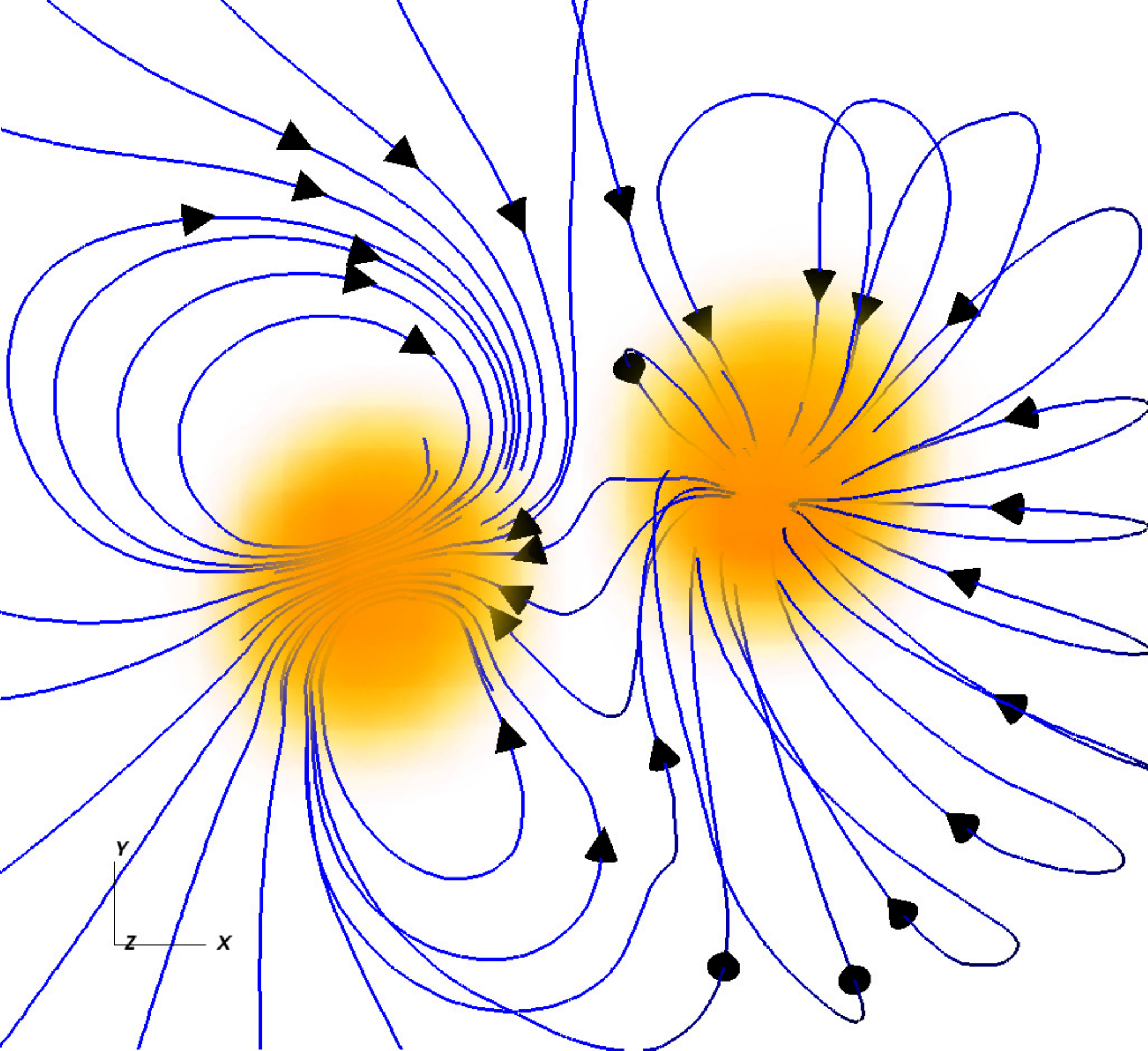}
        \includegraphics[width=.24\textwidth]{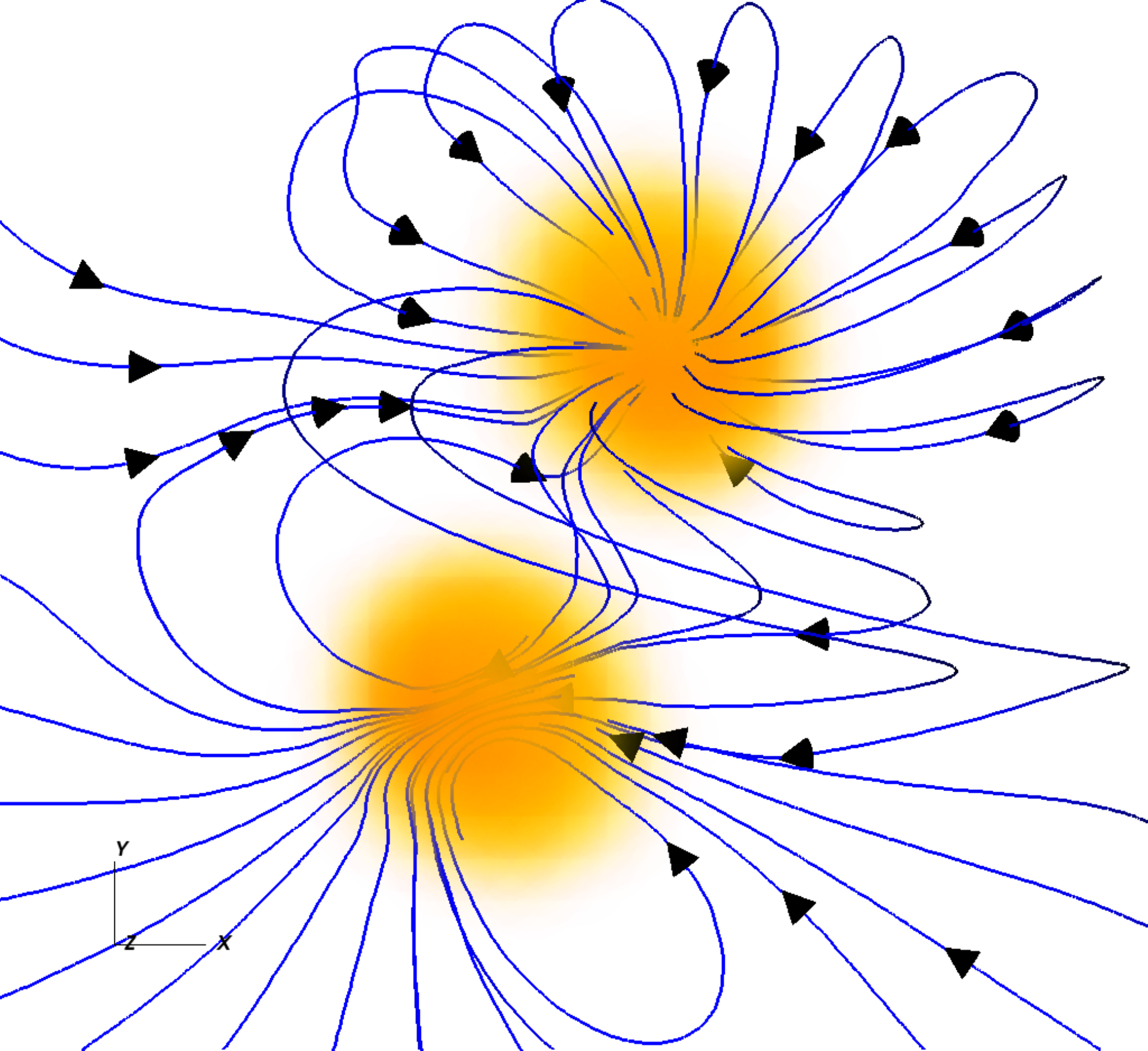}
        \\
        \vspace{-3.5mm}
        \hspace{-.22\textwidth}
        {\bf Aa)}\hspace{.21\textwidth}
        {\bf Ab)}\hspace{.21\textwidth}
        {\bf Ac)}\hspace{.21\textwidth}
        {\bf Ad)}
        \\
        \includegraphics[width=.24\textwidth]{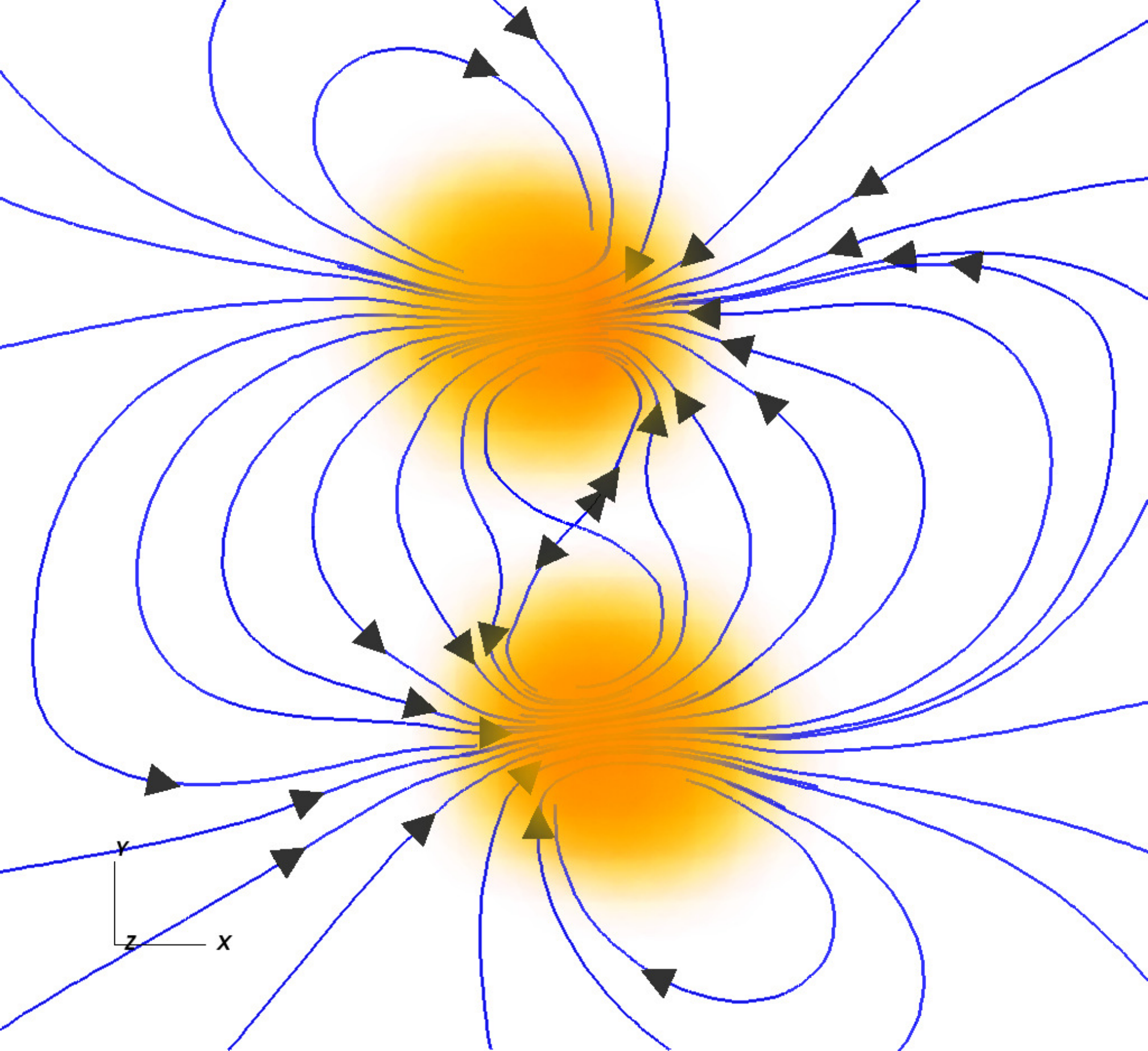}
        \includegraphics[width=.24\textwidth]{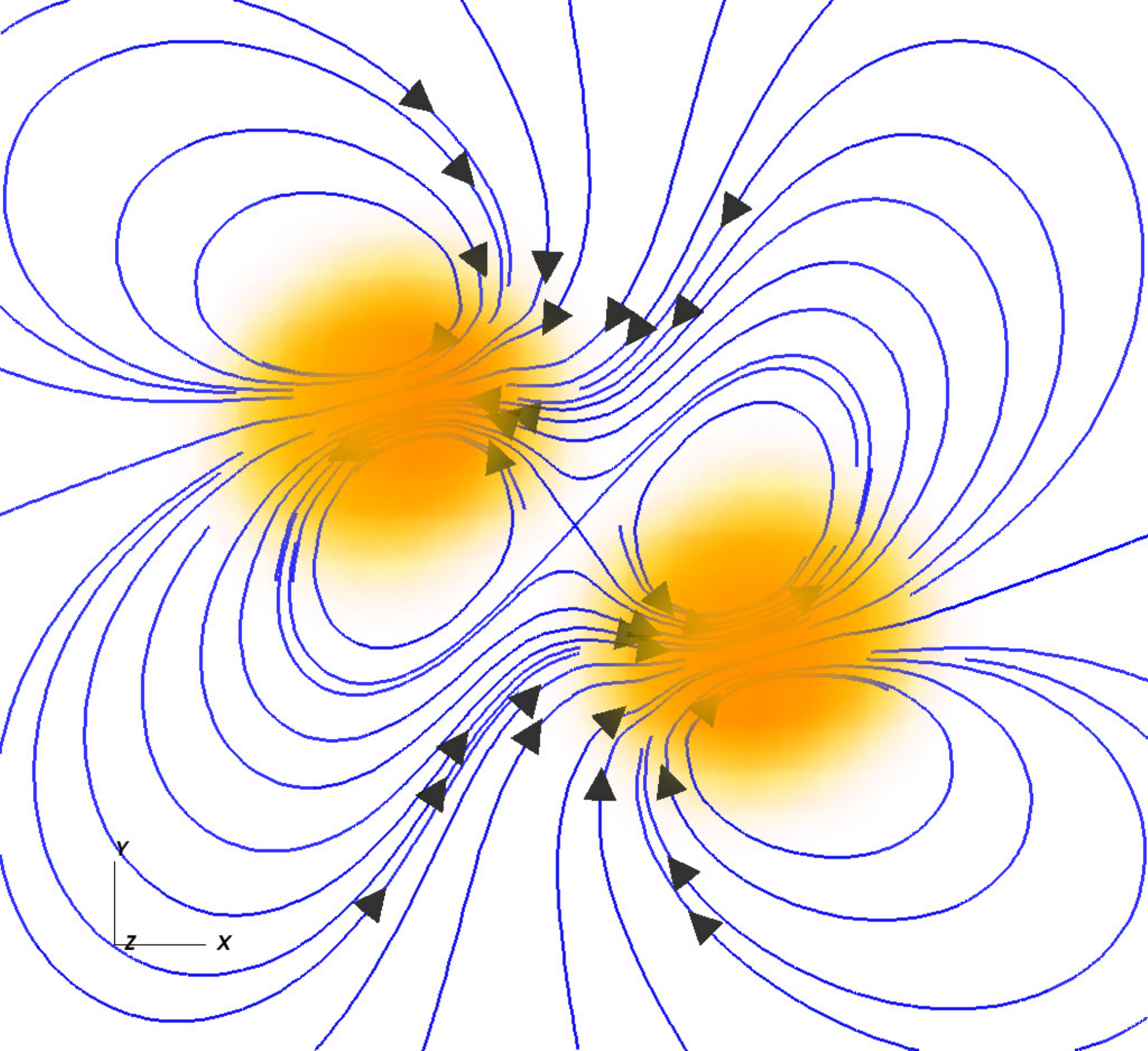}
        \includegraphics[width=.24\textwidth]{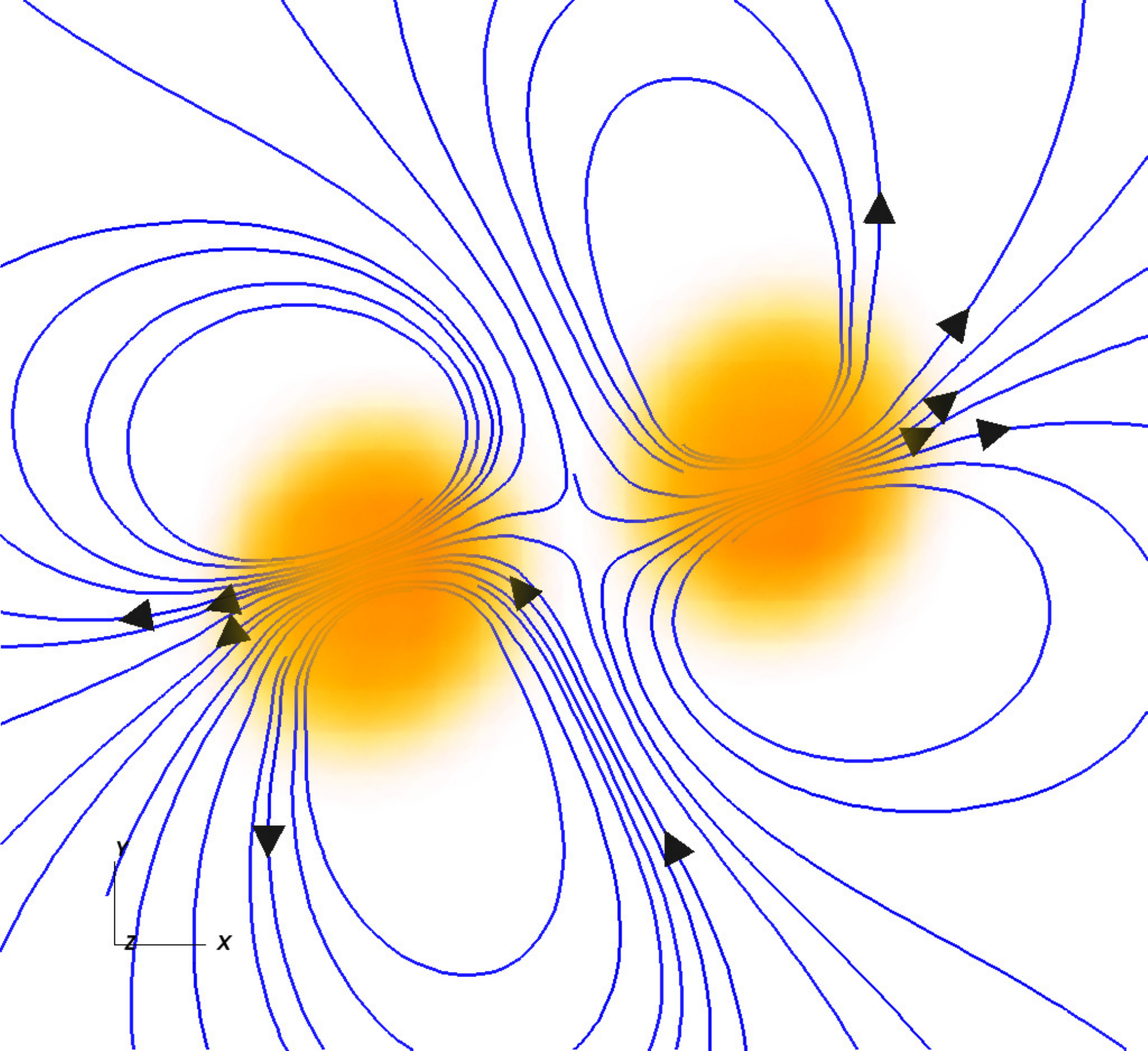}
        \includegraphics[width=.24\textwidth]{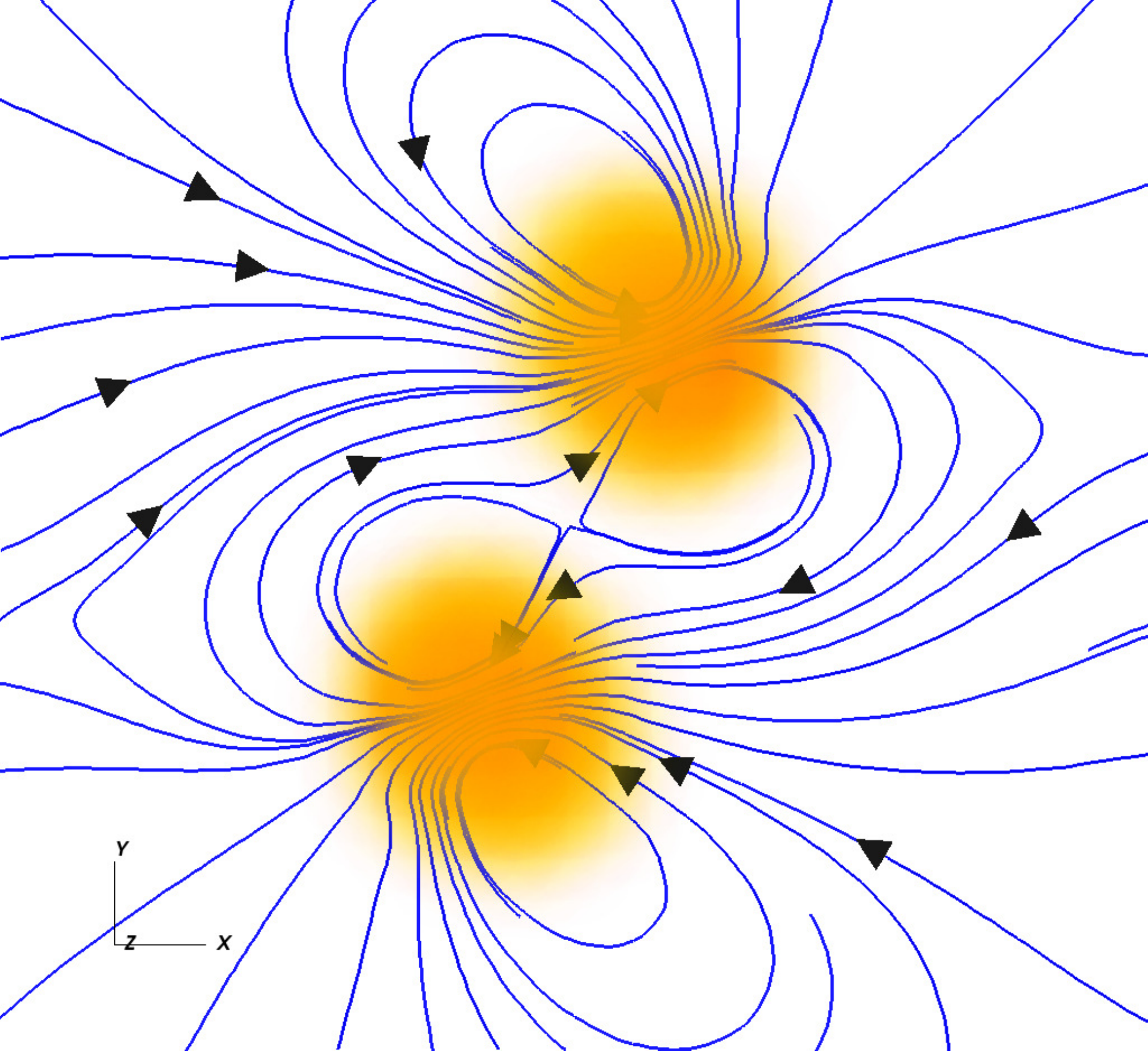}
        \\
	\vspace{-3.5mm}
        \hspace{-.22\textwidth}
        {\bf Ba)}\hspace{.21\textwidth}
        {\bf Bb)}\hspace{.21\textwidth}
        {\bf Bc)}\hspace{.21\textwidth}
        {\bf Bd)}
        \\
        \includegraphics[width=.24\textwidth]{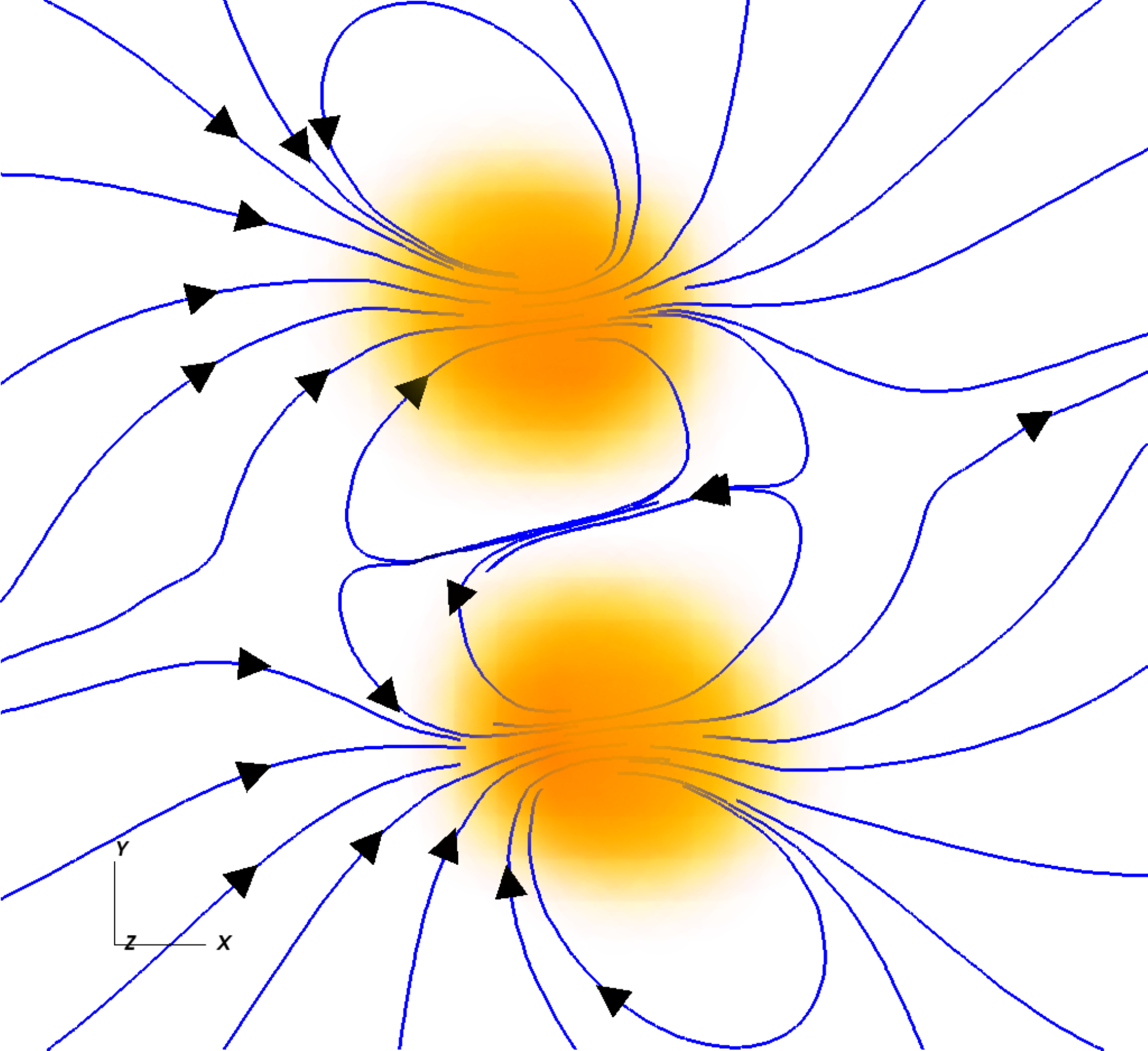}
        \includegraphics[width=.24\textwidth]{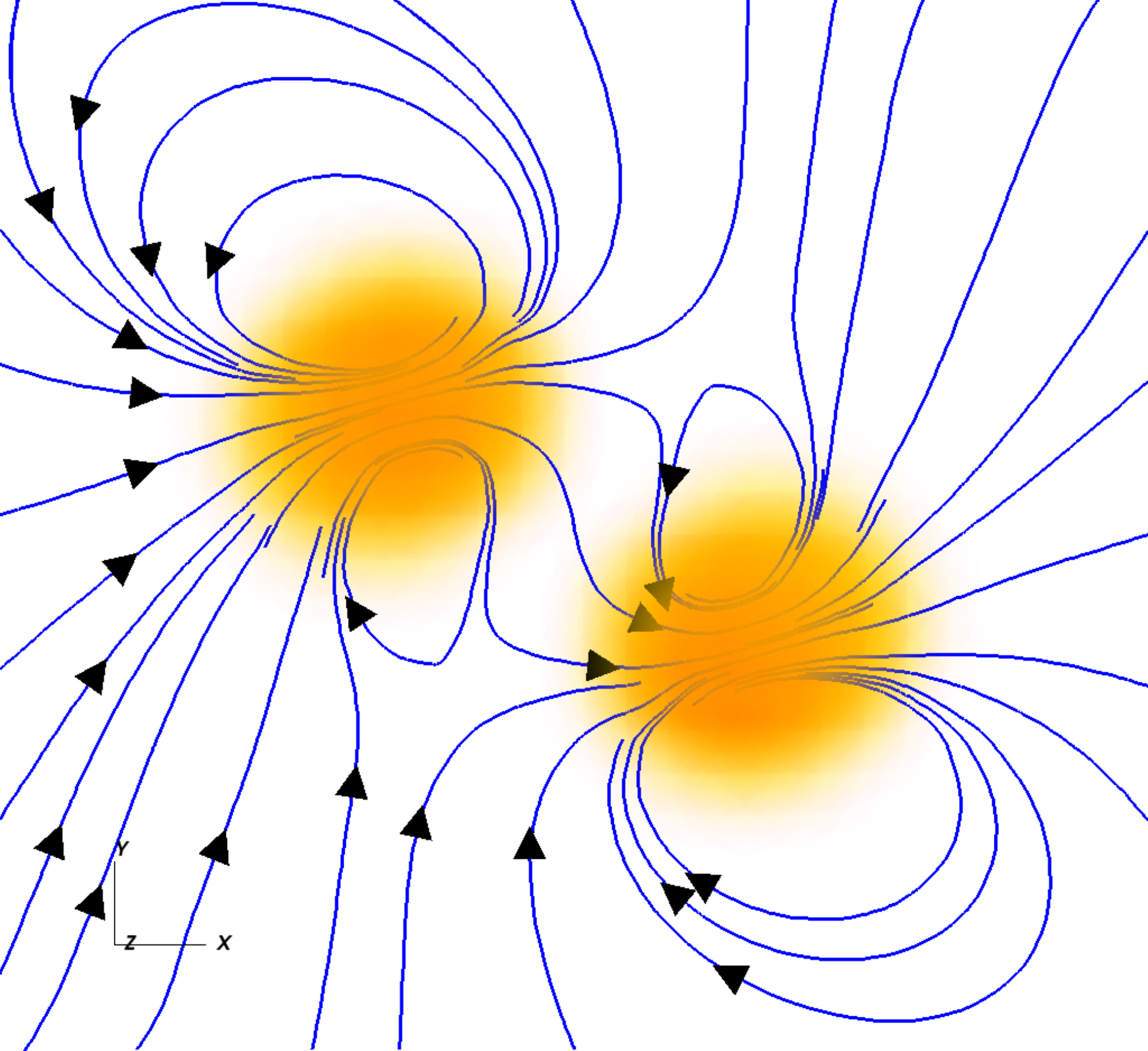}
        \includegraphics[width=.24\textwidth]{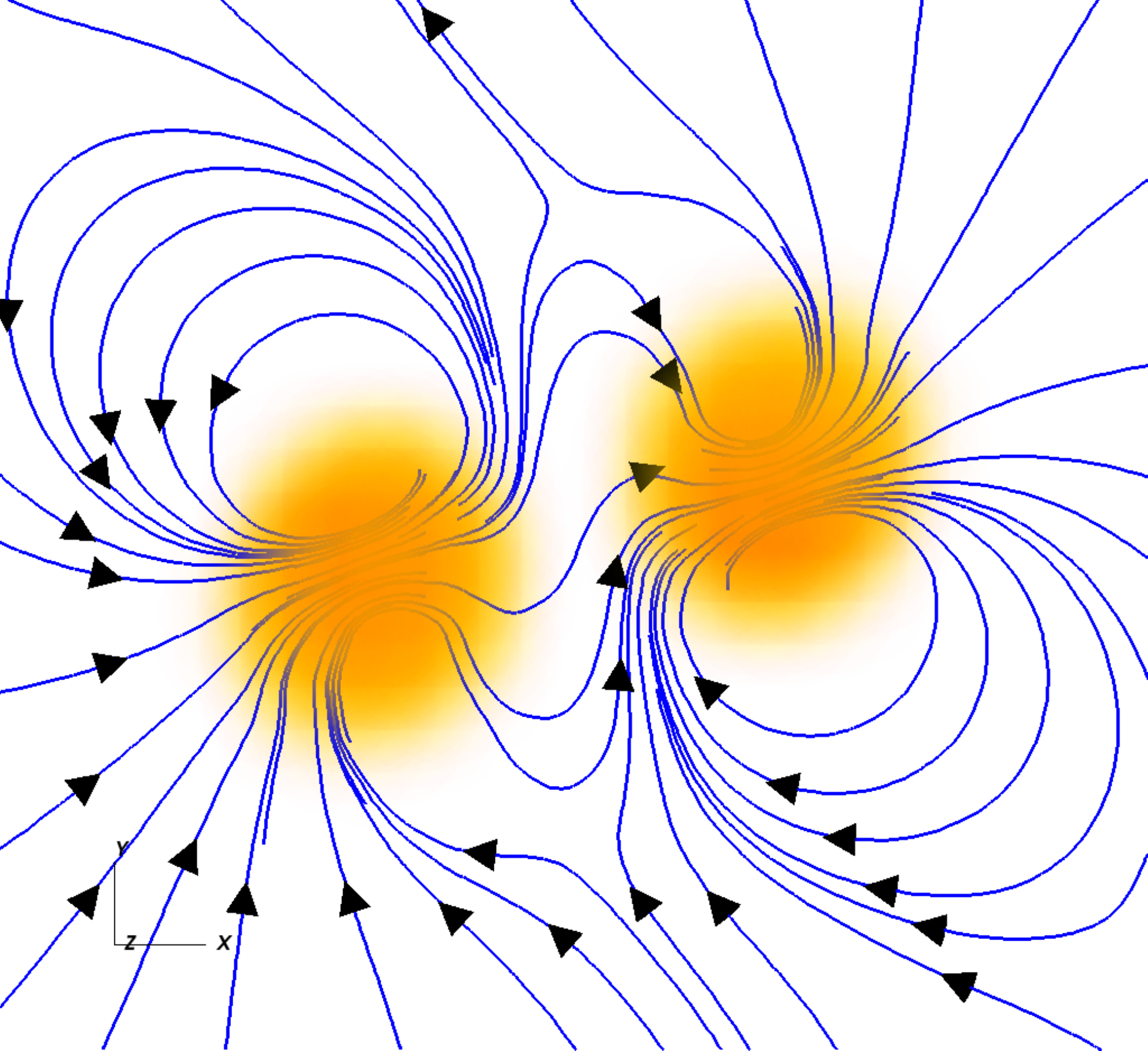}
        \includegraphics[width=.24\textwidth]{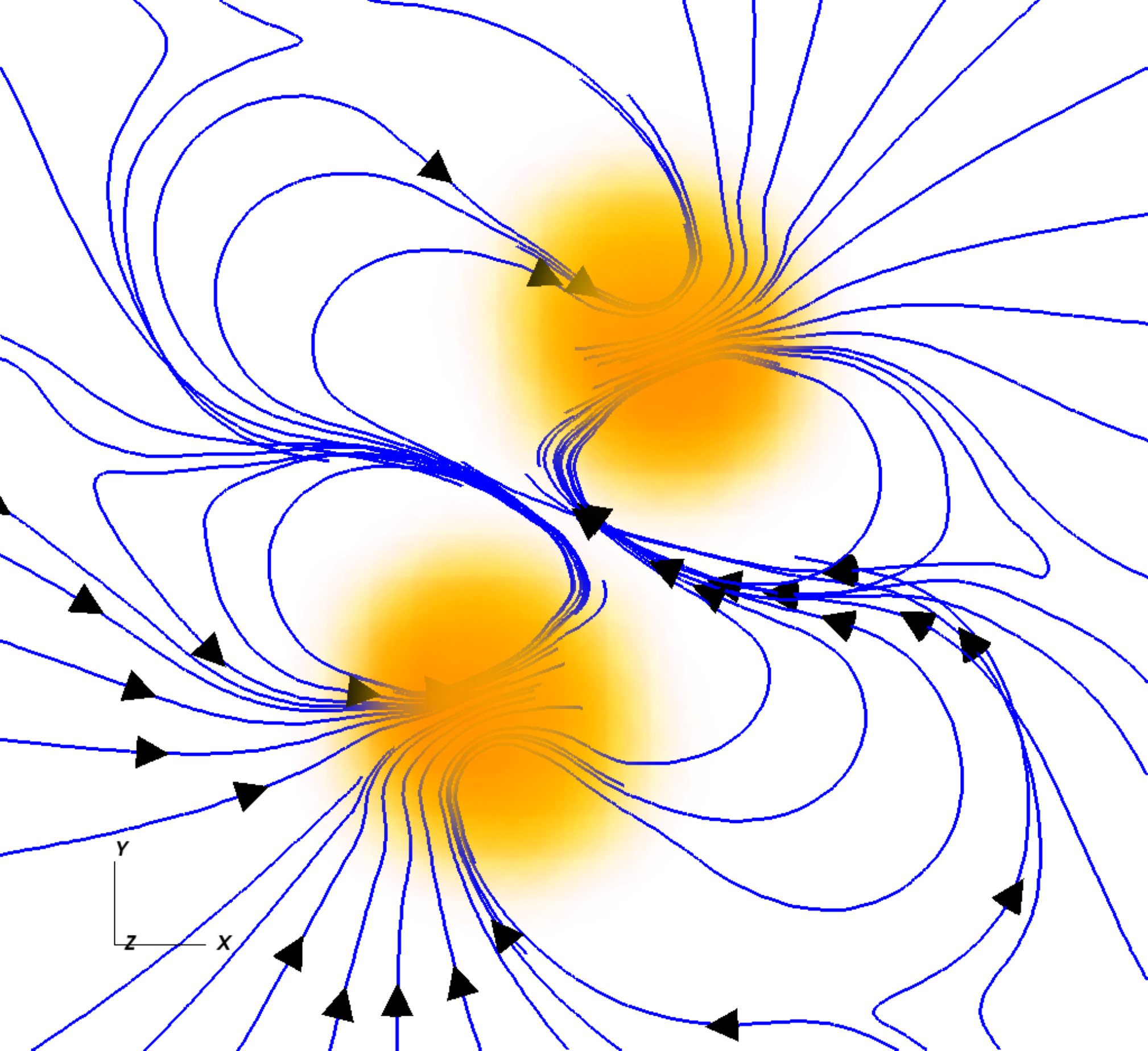}
        \\
	\vspace{-3.5mm}
        \hspace{-.22\textwidth}
        {\bf Ca)}\hspace{.21\textwidth}
        {\bf Cb)}\hspace{.21\textwidth}
        {\bf Cc)}\hspace{.21\textwidth}
        {\bf Cd)}
        \\
        \includegraphics[width=.24\textwidth]{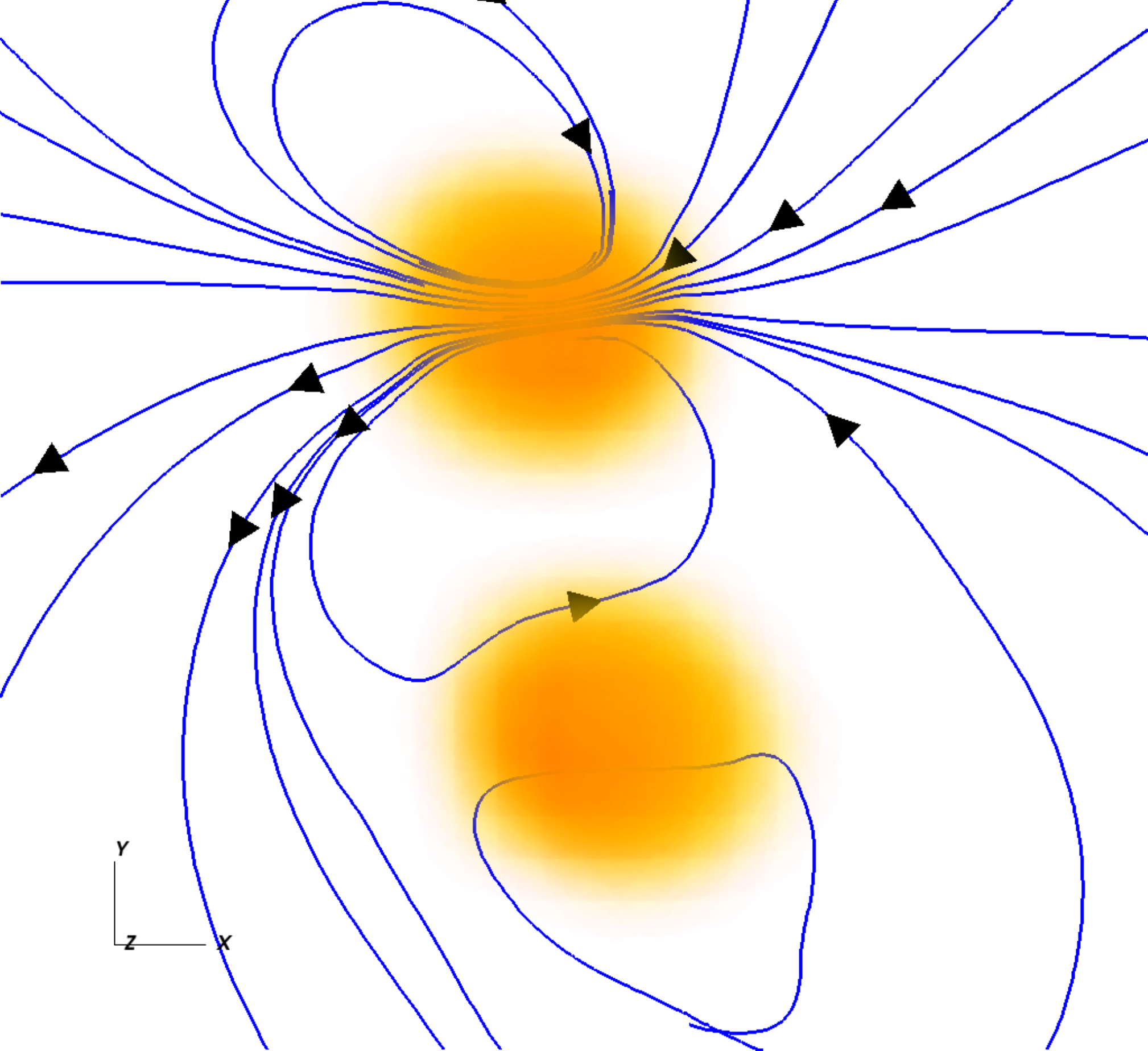}
        \includegraphics[width=.24\textwidth]{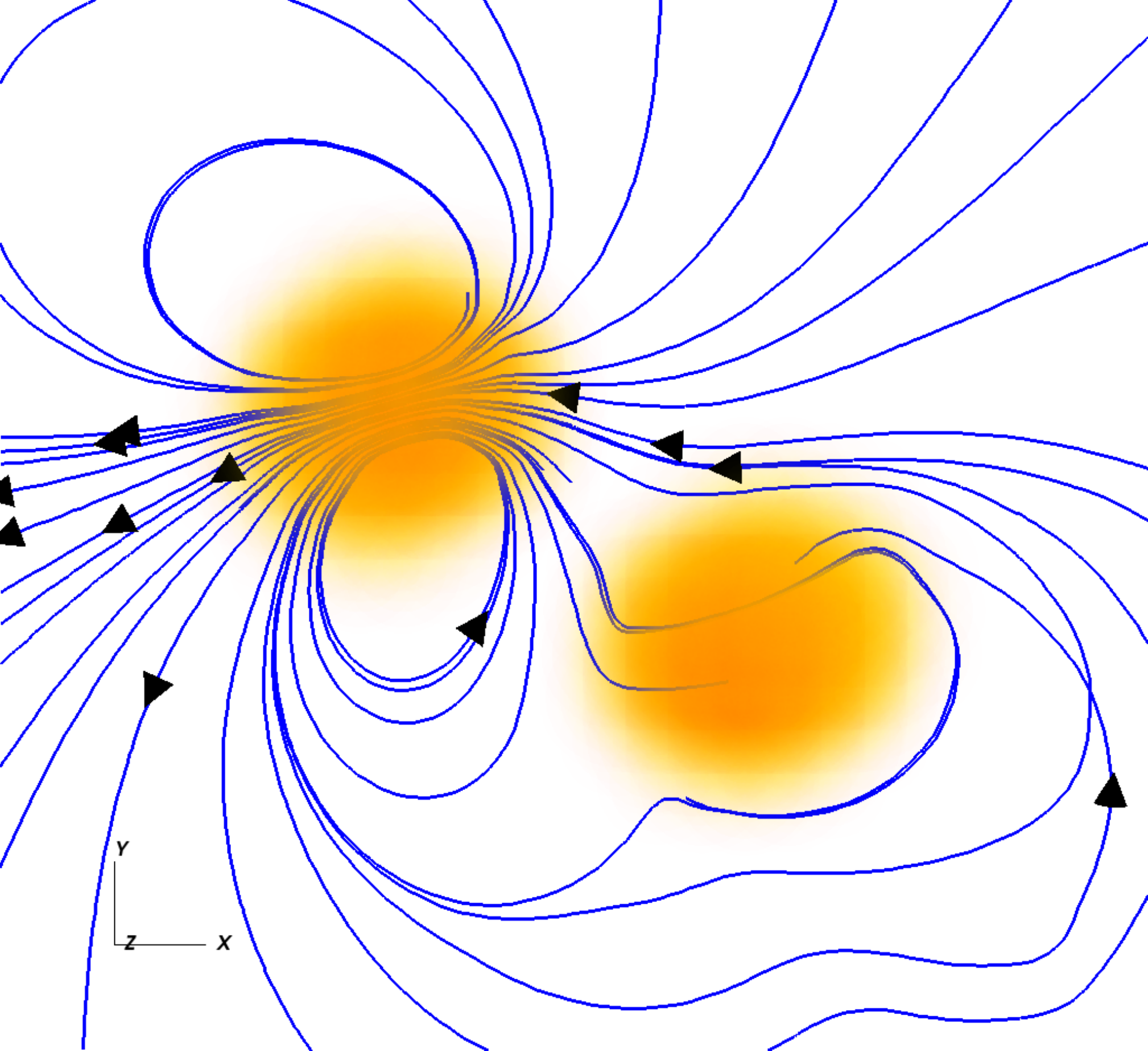}
        \includegraphics[width=.24\textwidth]{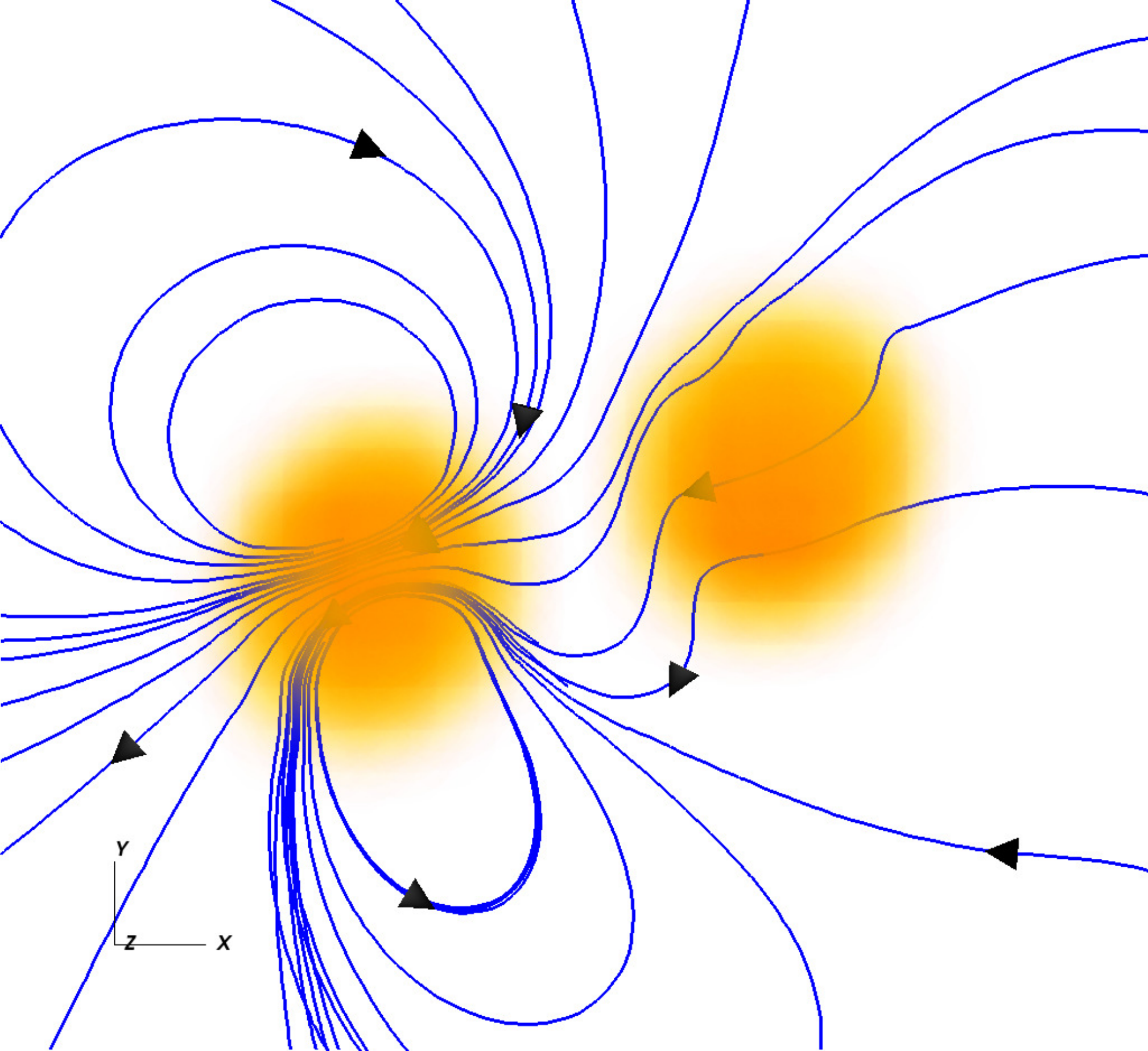}
        \includegraphics[width=.24\textwidth]{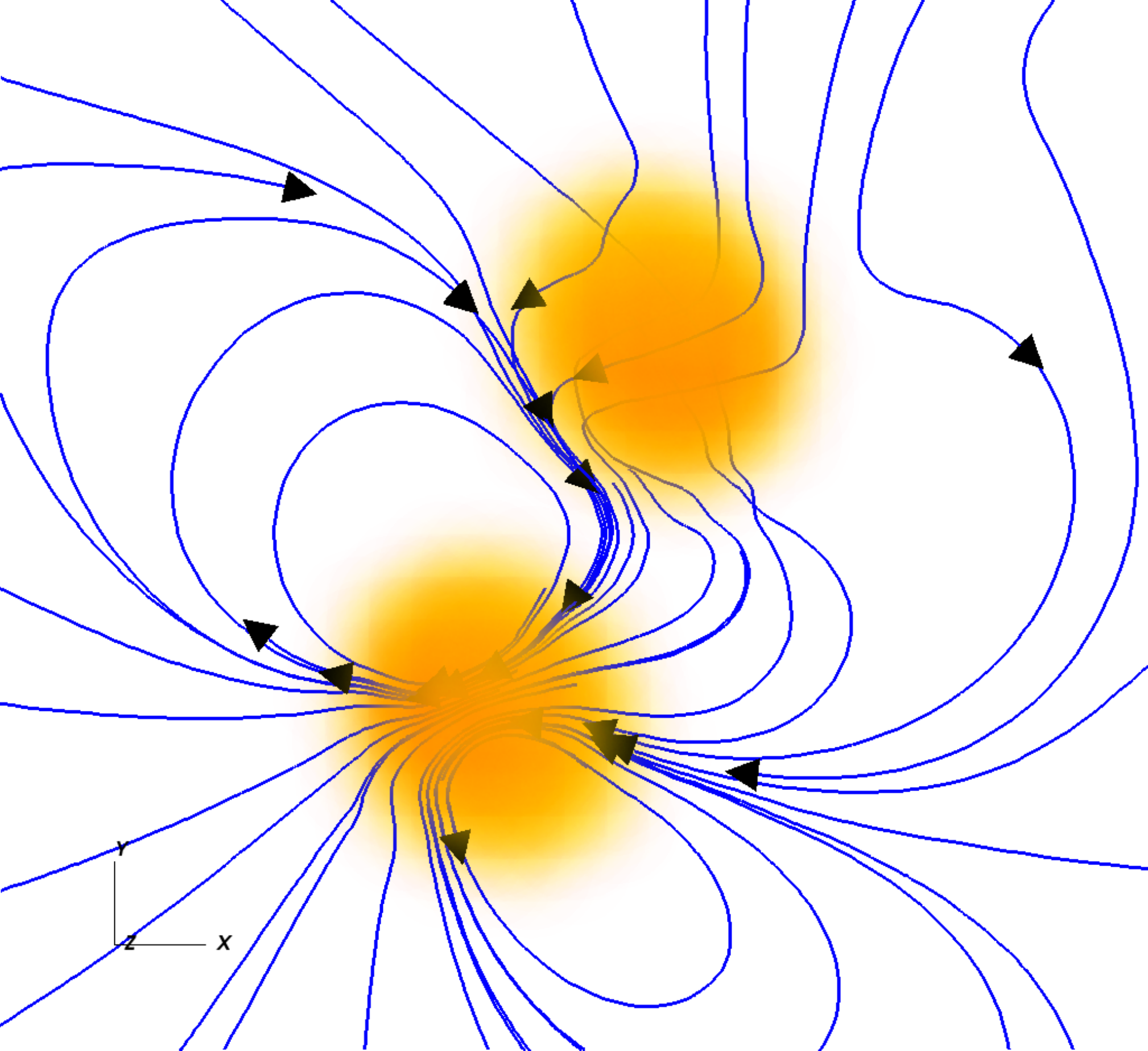}
        \\
        \vspace{-3.5mm}
        \hspace{-.22\textwidth}
        {\bf Da)}\hspace{.21\textwidth}
        {\bf Db)}\hspace{.21\textwidth}
        {\bf Dc)}\hspace{.21\textwidth}
        {\bf Dd)}
        \\

        \caption{
		Stellar density (yellow) and magnetic field lines (blue) seeded on the orbital plane.
		The columns { {\bf a}, {\bf b}, {\bf c}, {\bf d}} display the configurations
		at times $t \approx \{-3.6, -3.2, -2.8, -2.2\} \rm{ms}$, where
		time intervals between successive columns describe roughly one eighth of an orbit.
                The radius of each star is $R=13.6 {\rm km}$ and the binary initial separation 
                is $a=45 {\rm km}$.
                {\bf Row A:} magnetic dipoles perpendicular/parallel to the orbital ($P/U$),
                {\bf Row B:} magnetic dipoles parallel to the orbital plane (opposite orientations) (\textit{P/-P}), and
                {\bf Row C:} magnetic dipoles parallel to the orbital plane (same orientations) ($P/P$), and
                {\bf Row D:} non-magnetized/magnetized stars with magnetic dipole in the orbital plane ($P/u$).
	Notice that field lines depicted in Row A appear to cross other lines, but they actually leave the orbital plane
     and do not cross (see Fig.~\ref{fig:3d-Bfields}).
}
        \label{fig:Bfields}
\end{figure*}

%---------------------------------------

\begin{figure}
	\includegraphics[width=.75\columnwidth]{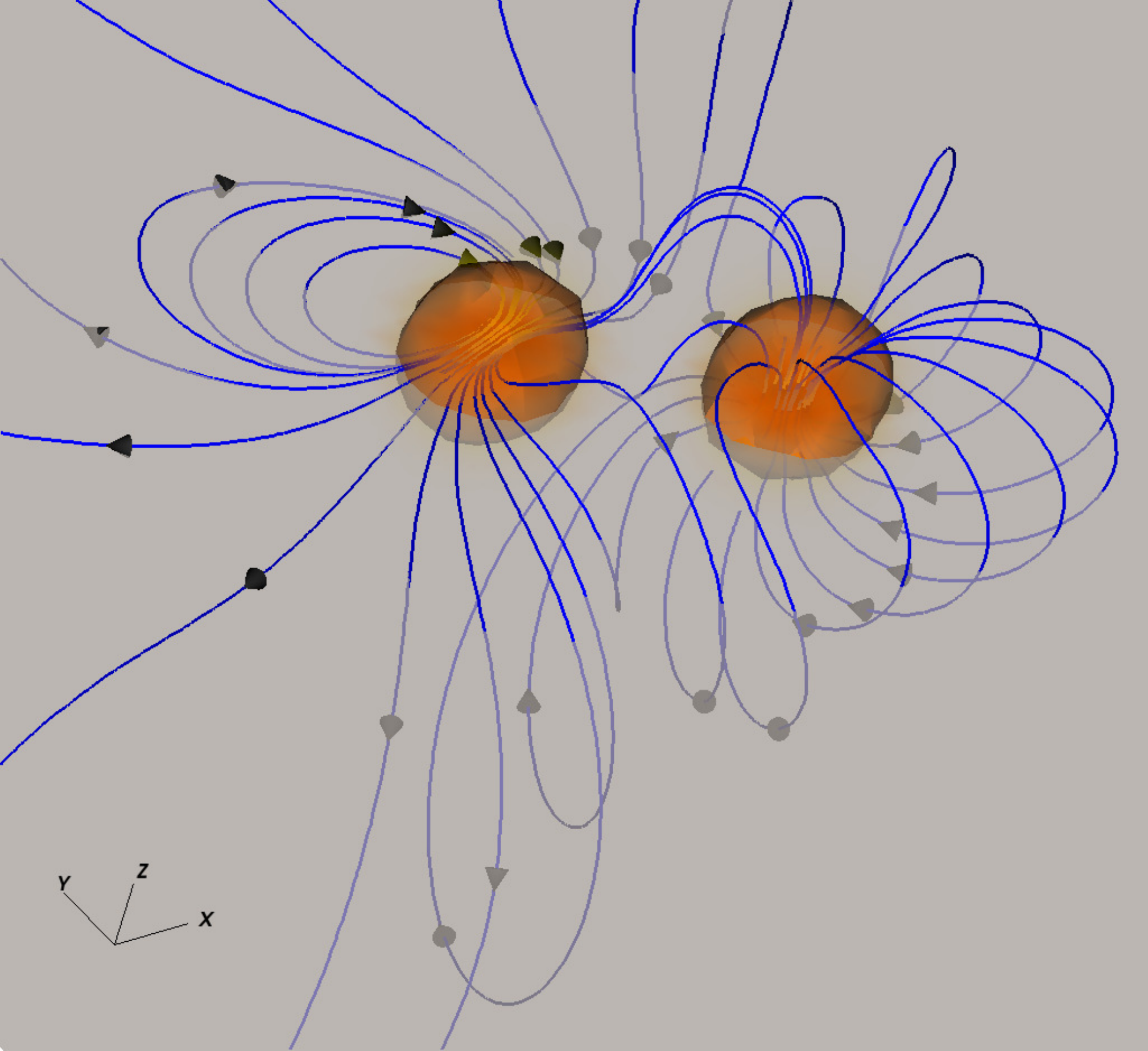}
	\caption{Three-dimensional arrangement of field lines for the $P/U$ case, roughly $3 {\rm ms}$ before merger.
        We have made the orbital plane translucent (displayed in grey) 
        to aid in distinguishing sections of the field lines lying above from those
        below the $z=0$ plane (the latter darker in color).
        Notice this case has field lines that naturally venture  off the orbital plane.
         }\label{fig:3d-Bfields}
\end{figure}

%---------------------------------------

\begin{figure*}
	\centering
        \includegraphics[width=0.95\columnwidth]{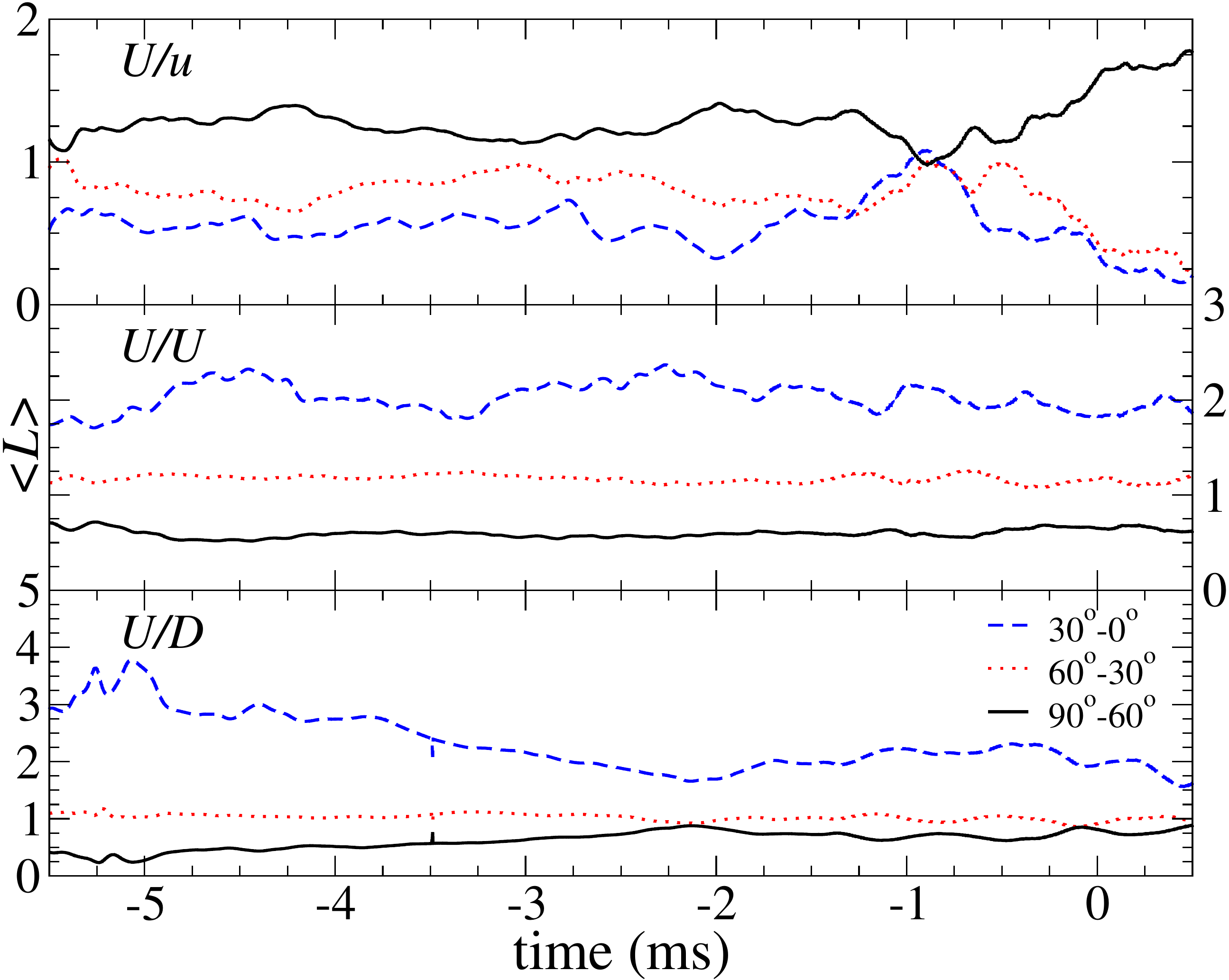}
        \includegraphics[width=0.99\columnwidth]{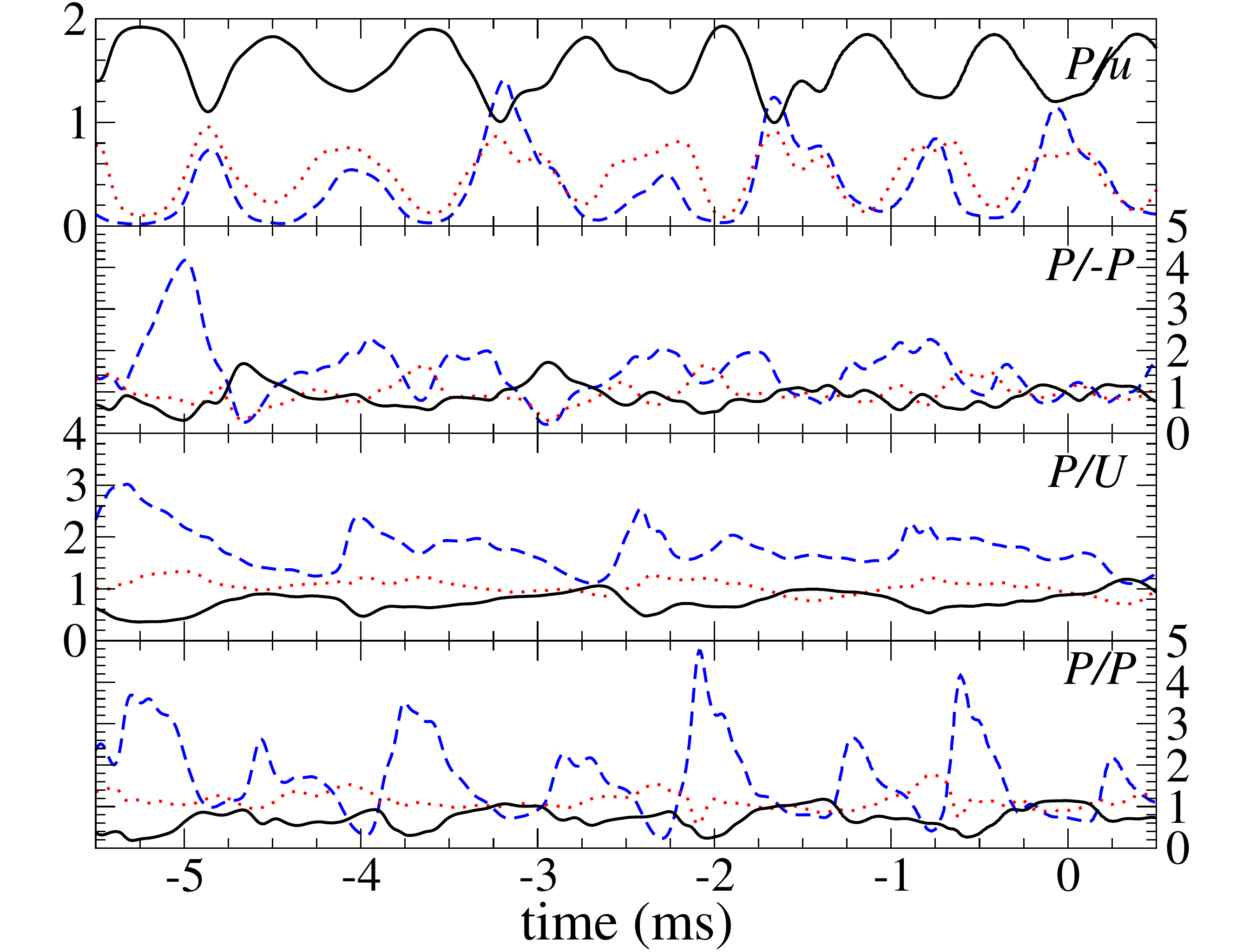}
        \caption{\textit{Left-panel}: Average luminosity per solid angle
        $\langle L \rangle
         \equiv L_{\langle\theta_1,\theta_2 \rangle}/[2\pi(\cos\theta_2-\cos\theta_1)]$,
        normalized by the total luminosity density $L_{T}/(4\pi)$,
        for the three aligned cases:
        $U/U$ (first row), $U/u$ (second row) and $U/D$
        (bottom row). Notice that in the $U/U$ and $U/D$ cases, the radiation
        patterns appear collimated, while the $U/u$ case radiates mostly close to
        the orbital plane of the binary.
	\textit{Right-panel}: Average luminosity per solid angle for the four
        misaligned cases: $P/u$ (first row), $P/U$ (second row), \textit{P/-P} (third row)
        and $P/P$ (bottom row). The radiative patterns of these configurations are more
        dynamic than those in the left panel (see also Fig.~\ref{fig:sqphi2Spheres}). 
	}\label{fig:lumsAngle}
\end{figure*}

%---------------------------------------

\begin{figure}
        \includegraphics[width=.70\columnwidth]{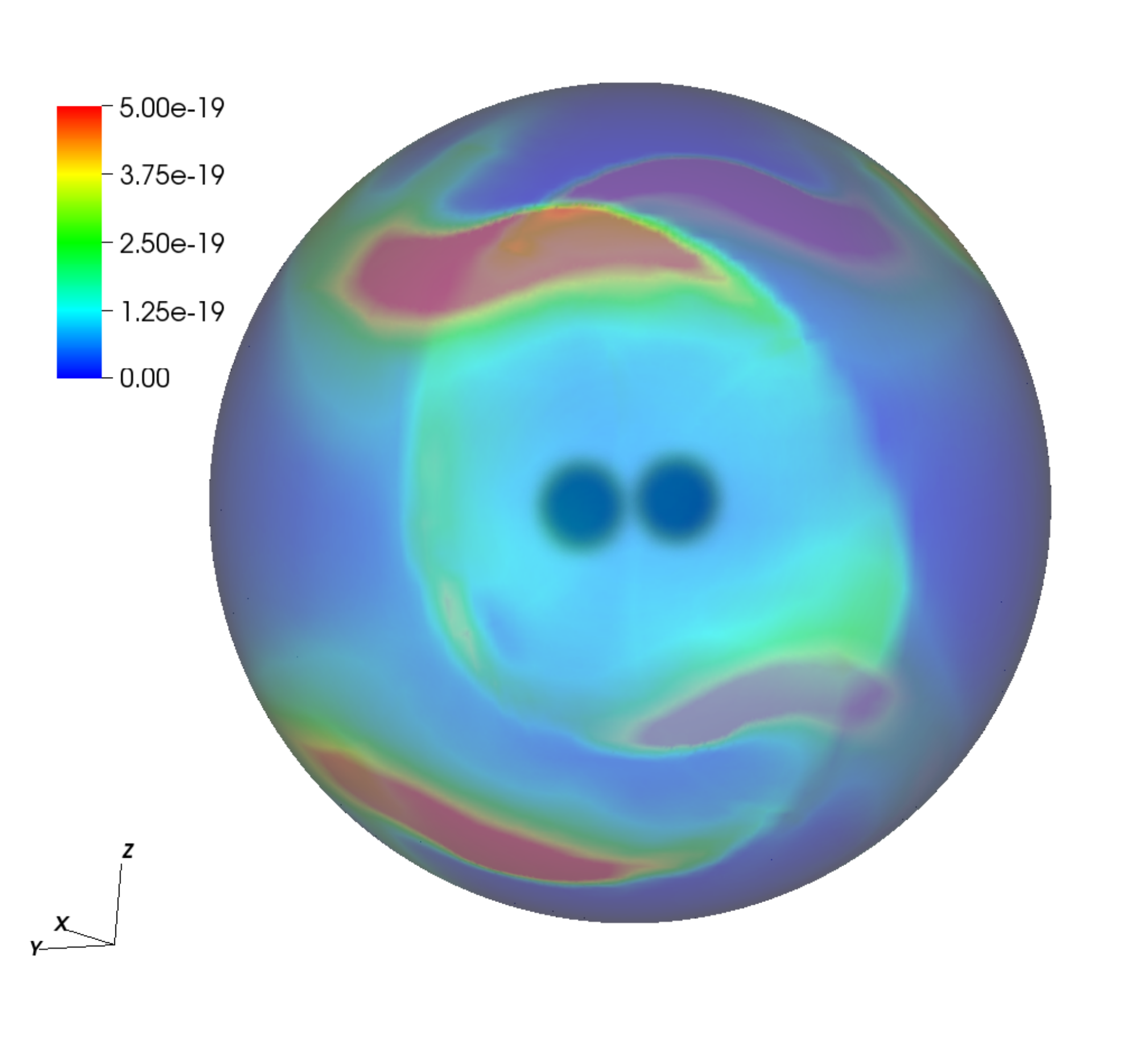}
        \includegraphics[width=.70\columnwidth]{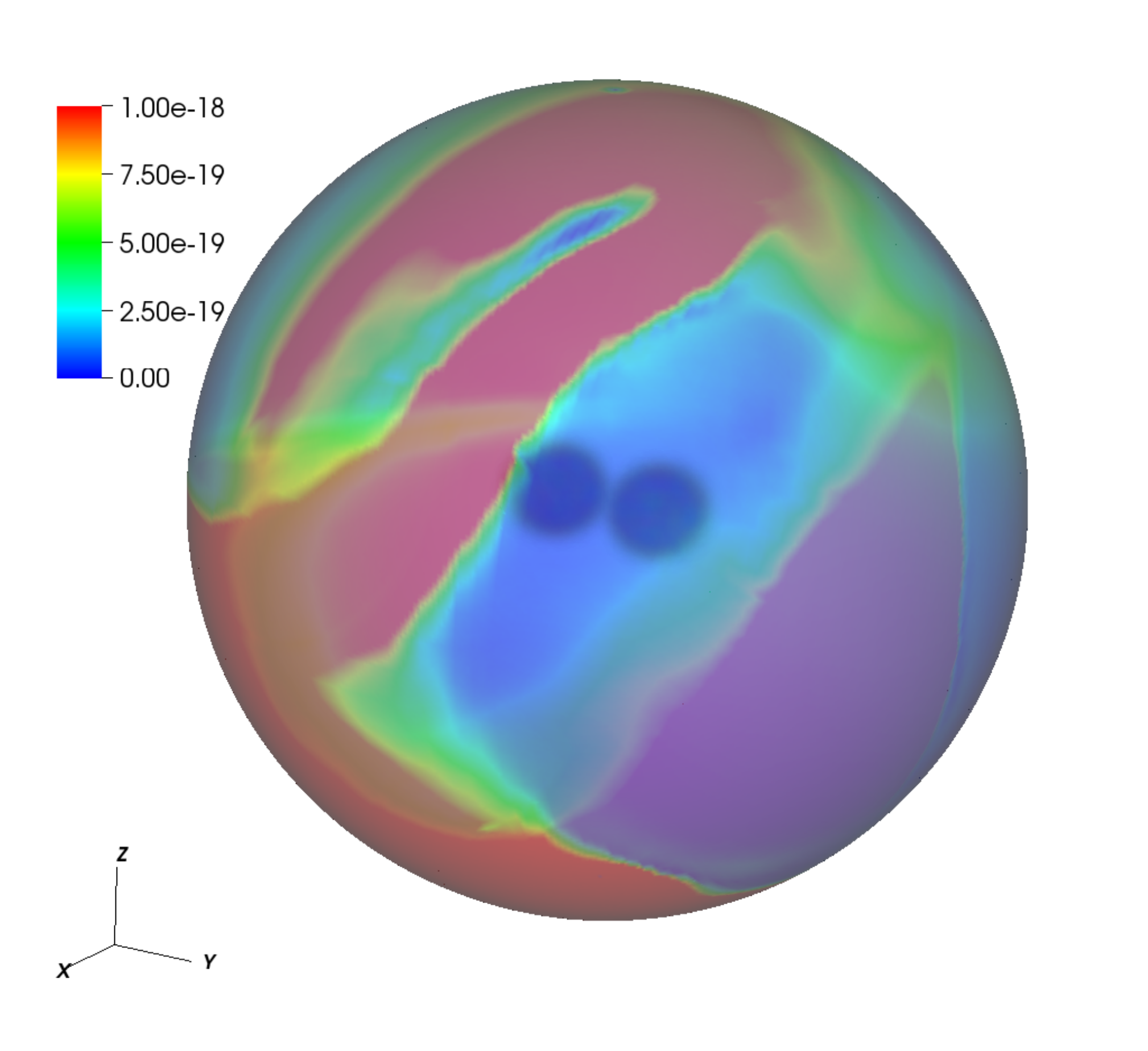}
        \includegraphics[width=.70\columnwidth]{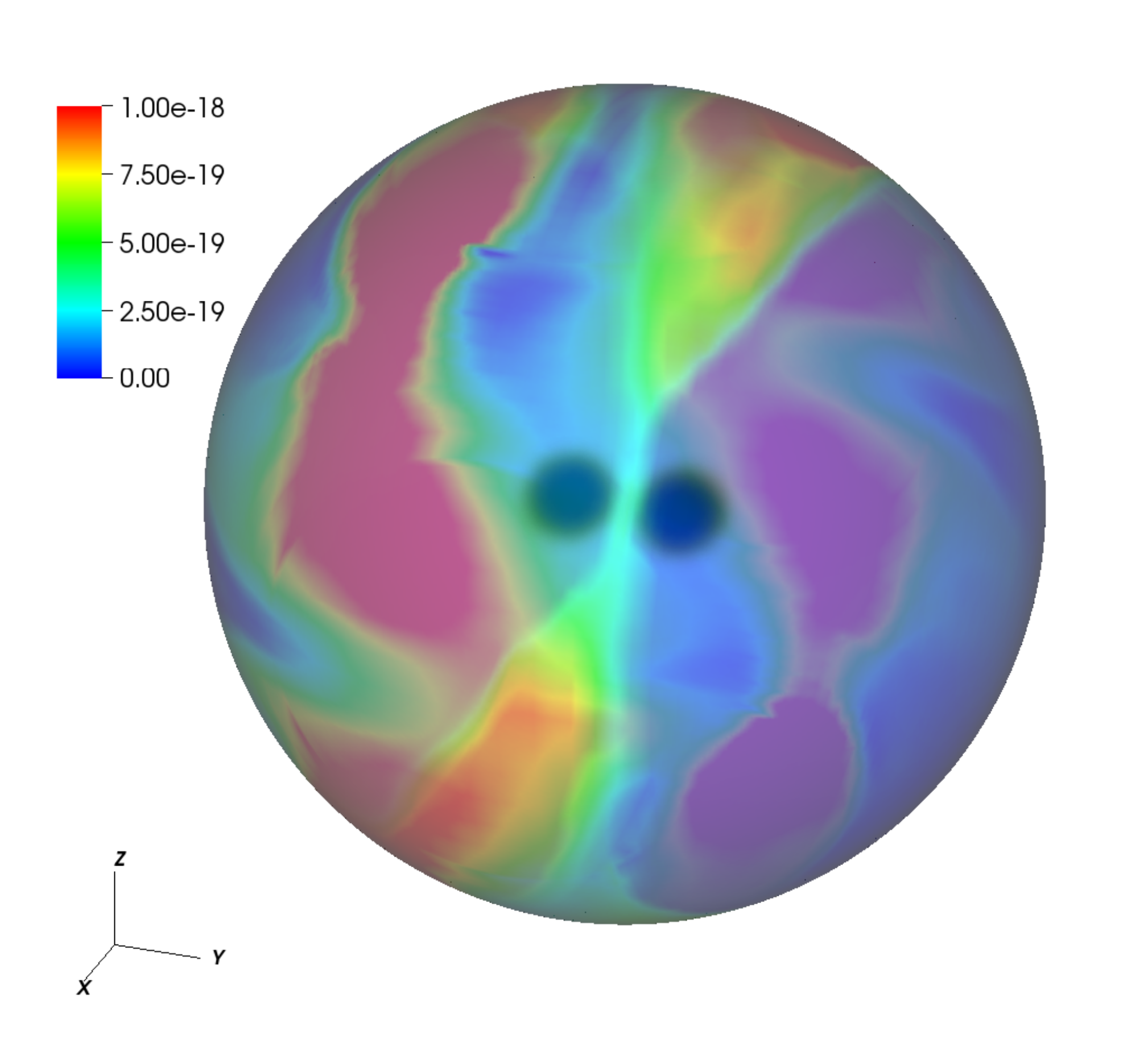}
        \includegraphics[width=.70\columnwidth]{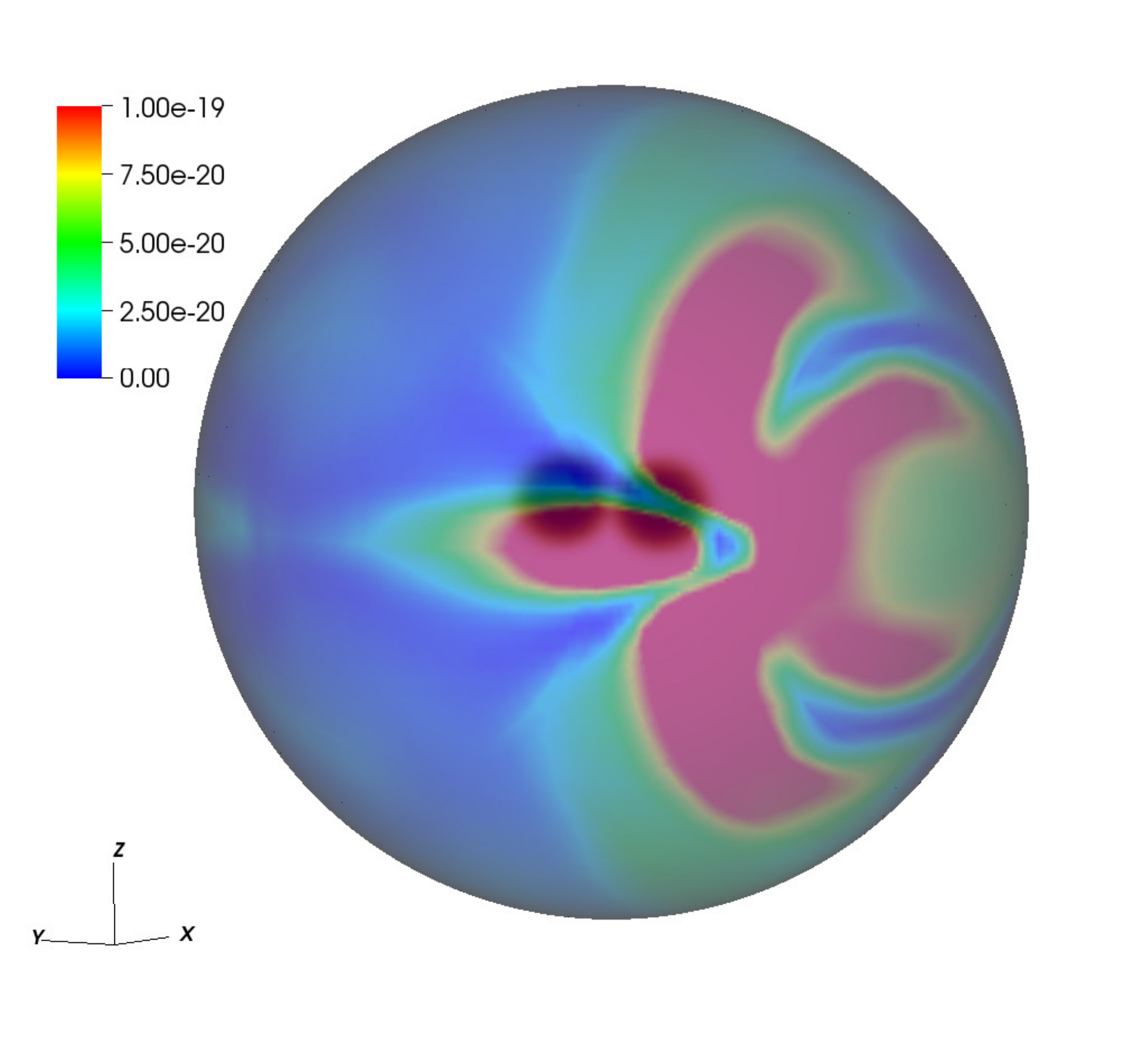}
        \caption{Poynting flux evaluated on an encompassing sphere at approximately 120km from the
         binary, roughly $t \approx 1.5$ms before merger.
        The \textit{P/P} (first row), \textit{P/U} (second row), \textit{P/-P} (third row), and \textit{P/u} (fourth row)
        are shown from top to bottom ---in decreasing order of collimation.
        Notice however, that the more highly collimated emissions are episodic
        (as shown in Fig.\ref{fig:lumsAngle}).}
        \label{fig:sqphi2Spheres}
\end{figure}

%---------------------------------------

We now turn our attention to binary neutron stars. For simplicity
we have adopted a setup almost identical to the one described
in \cite{2013PhRvL.111f1105P,2013PhRvD..88d3011P},
except for the orientation of the magnetic dipole moments of the stars.
We thus consider a binary of identical, irrotational
neutron stars with baryonic mass $M=1.62 M_{\odot}$ and radius
$R_*=R_c=13.6 {\rm km}$. The binary is initially in a quasicircular orbit with
angular velocity $\Omega_o=1.85 {\rm rad}/{\rm ms}$ and
separation $a=45 {\rm km}$.
Consistent initial geometric and matter configurations for this system are obtained
with the LORENE library~\cite{lorene}, which adopts a polytropic equation of
state $P=K \rho^{\Gamma}$ with $\Gamma=2$ and $K=123 G^3 M_\odot^2/c^6$ (notice
that we use units with $G=c=M_\odot=1$). The binaries merge at $t \approx 7$ms,
and they are evolved for $2-3$ms afterwards. In what follows, for concreteness, we shift
$t=0$ to the time that the stars touch. During
the evolution, the fluid is modeled with an ideal gas equation of state
which allows for the formation of shocks.

The magnetic moment $\mu_i$ describes a dipolar magnetic field $B^i$ in
the comoving frame of each star. In our previous work~\cite{2013PhRvL.111f1105P,2013PhRvD..88d3011P}
we have considered scenarios in which the magnetic moment is aligned with
the orbital angular momentum, i.e. $\vec \mu=\mu \hat k$ (with $k$ the unit
vector along the $z$ axis and parallel to the angular momentum). Here we consider
less idealized cases in which the magnetic moment can be perpendicular
to the orbital angular momentum, i.e. $\vec \mu=\mu \hat i$ (with $i$ the unit
vector along the $x$ axis on the orbital plane).
Notice that the radial magnetic field at the pole
of the star $B_*$ is related to the magnitude of the magnetic moment
by the relation $\mu = B_* R_*^3$. In all simulations considered, we adopt
$B_* = 1.5 \times 10^{11} {\rm G}$, a value which, while admittedly high, is within
the expected range for realistic neutron stars in binaries. 
The initial electric field is obtained
from the ideal MHD condition. 

We consider a range of configurations
with at least one of the stars containing a magnetic moment perpendicular to the
orbital angular momentum:
\begin{itemize}
\item $P/u$: one-dominant, perpendicular case $\mu^{(1)}_x=100\, \mu^{(2)}_z=\mu$,
       where one of the magnetic moments is aligned with the orbital angular
       momentum but having a much smaller magnetization than the perpendicular one.
\item $P/U$: perpendicular-aligned case $\mu^{(1)}_x= \mu^{(2)}_z = \mu$, where
       one of the magnetic moments is aligned with the orbital angular momentum
       and the other is perpendicular to it.
\item \textit{P/-P}: perpendicular-antiperpendicular $\mu^{(1)}_{x} = -\mu^{(2)}_{x}$,
      where both magnetic moments are anti-parallel to each other and perpendicular to
      the orbital angular momentum.
\item $P/P$: perpendicular-perpendicular $\mu^{(1)}_{x} = \mu^{(2)}_{x}$, 
      where both magnetic moments are parallel to each other and perpendicular to
      the orbital angular momentum.
\end{itemize}

We recall that in our previous study we have already considered the following three cases
with magnetic moments parallel to the orbital angular momentum:
\begin{itemize}
\item $U/U$: aligned case $\mu^{(1)}_z=\mu^{(2)}_z=\mu$,
\item $U/D$: anti-aligned case  $\mu^{(1)}_z=-\mu^{(2)}_z=\mu$,
\item $U/u$: one-dominant, aligned case  $\mu^{(1)}_z=100\, \mu^{(2)}_z=\mu$.
\end{itemize}
For the rest of the paper, we refer collectively to these three  cases as
the {\em aligned} cases for simplicity because both magnetic
dipoles are perpendicular to the orbital plane, despite the fact
that the $U/D$ case is anti-aligned with the orbital angular momentum.

Our numerical domain extends to $320~{\rm km}$ in each direction
and contains five,
centered FMR grids with decreasing side-lengths (and twice as well resolved)
such that the highest resolution grid has $\Delta x = 300~{\rm m}$
and extends up to $58~{\rm km}$, covering both stars and the inner
part of the magnetosphere. Within this setup, we study the four cases
described above: $P/u$, $P/U$, \textit{P/-P}, and $P/P$ and compare with
those previously studied $U/U$, $U/D$ and $U/u$.

Of particular relevance is the behavior
of the Poynting flux resulting from the dynamics and
its dependence on the magnetic dipole orientation.
The luminosities, displayed as a function of time in Fig.~\ref{fig:lums},
are computed by integrating the outgoing Poynting flux on a surface located
at $R_{\rm ext}= 180~{\rm km}$. We also extract this quantities at two further
radii to confirm the expected $r^{-2}$ asymptotic behavior of the Poynting flux.
As evident in the figure, the total luminosity differs
across all cases considered; not only does the strength vary depending
on orientation but also temporal modulations are induced. The luminosities
are roughly periodic, accompanied by an overall increase 
as the orbit tightens.

As we discuss in more detail below, these temporal
modulations appear because of complicated dynamics between the stellar
magnetospheres. In particular, our results indicate that the 
local maxima (peaks) in the luminosity arise, for both stars equally magnetized (i.e.
the cases $P/U$, $P/P$, \textit{P/-P}), at the times when the magnetic fields emanating from each star
reconnect.
In other words, these three cases are roughly cyclic, and at one point in the cycle,
the field lines from one star reconnect with those of the other, causing a peak
in the luminosity.
Subsequent to the reconnection, the lines elongate until they must
reconfigure to begin a new cycle.

The magnetic field configurations and the stellar
density are shown at different representative times in Fig.~\ref{fig:Bfields}.
In order to illustrate the relationship between the field dynamics and the oscillations
in the luminosity, we identify particular physical
events, such as when the stars have severely stretched their connecting field lines 
or when reconnection between the stars reconfigures those connecting lines.
By accounting for the propagation time to the extraction surface,
we establish certain correlations for each case as described below.

We begin with the $P/U$ case, which is the most complex one as 
in that the bilateral
symmetry ($z\rightarrow -z$) is broken and the field lines cross the orbital plane
(see Fig.~\ref{fig:3d-Bfields}). In this case,
strong reconnections arise when the magnetic moment perpendicular to the
angular orbital momentum ($P$) points away from the other star 
(see Fig.~\ref{fig:Bfields}-Ac).  At such a stage, short field lines from
the $U$ star connect with the $P$ star as shown in Fig.~\ref{fig:3d-Bfields}.
These tight, linking field lines are subsequently stretched until reconnection
re-establishes the tight linking. This reconnection is thus expected to
occur with every half-orbit and 
such a periodicity is evident in the total luminosity for this case as shown
in Fig.~\ref{fig:lums}.

Let us turn now to the \textit{P/-P} case. Here, when the polar regions face each other, the field lines
from each star are oppositely-directed resulting in large deflections
(see Fig.~\ref{fig:Bfields}-Bc). A quarter of an orbit
 later---when the ``lateral'' sides of each star face each other---the field lines emanating from
one star towards the other can reconnect with those of the companion star (see Fig.~\ref{fig:Bfields}-Ba).
Here again, as the orbit proceeds, these lines stretch until they must reconfigure by reconnection 
to repeat the cycle. These reconnection events give rise to the observed local maxima,
which occur twice per orbit. 
The behavior exhibited by the \textit{P/P} case is in clear contrast to the \textit{P/-P} case. 
Namely, in this binary, it is when polar regions face each other  that
fields emanating from one star can reconnect with those entering the companion
(see Fig.~\ref{fig:Bfields}-Cc). As the orbit proceeds, the lines stretch and reconfigure
to repeat the cycle. Consequently the 
geometry of the  \textit{P/-P} case is such that the tightest configuration of field lines occurs a half-orbit
later than the corresponding tightest configuration of the $P/P$ case. This offset is also evident between the
local peaks in the two luminosities in Fig.~\ref{fig:lums}.

In addition to the oscillatory features, the overall luminosities of these two cases (\textit{P/-P} and $P/P$)
are roughly an order of magnitude different. Similar to the $U/U$ and $U/D$ cases, we attribute this difference
to the fact that the more luminous of the pair belongs to that with the ``tightest'' magnetic configuration, where
by ``tightest'' we roughly mean the minimum (electrovacuum) potential energy
$U=-\vec{\mu}_1 \cdot \vec{B}_{2|1} -\vec{\mu}_2 \cdot \vec{B}_{1|2}$
achieved by the two dipoles during their orbit (with $B_{i|j}$ denoting the
magnetic field induced by the star $i$ at the location of star $j$). Just as the $U/D$ case has less potential energy than the $U/U$ case and is
more luminous, we find here that the $P/P$ case is similarly more luminous than the \textit{P/-P} case.

The luminosity of the $P/u$ case also displays temporal modulations, but notably
the local maxima at early times seem correlated with those of the $P/P$ case. 
This correlation can
be understood by first noticing that for the separations considered,
the magnetic field in the neighborhood of the weakly magnetized star
is dominated by that from the strongly magnetized star.
As the $u$ star passes through the region near the polar caps of the strongly magnetized $P$ star,
the disruption to the ambient field is maximized (see for instance Fig.~\ref{fig:Bfields}-Dc)
and it is at these points in the 
orbit that the luminosity peaks occur. These points occur twice an orbit
and geometrically coincide with the configurations of the $P/P$ case which produce
peaks in its luminosity.

Interestingly, all obtained luminosities are roughly within the
range defined by our previously studied aligned cases \cite{2013PhRvL.111f1105P,2013PhRvD..88d3011P};
i.e. the $U/u$ and $U/D$ cases represent lower and upper limits. 
In addition,
those previous cases did not show any significant temporal modulation, which arises
here as a result of the misalignment of magnetic moments with respect to the
orbital angular momentum.
More importantly, inspection of the luminosity per solid angle for the misaligned cases
considered here indicates only an episodic collimation of the resulting Poynting flux
for all cases (see Fig.~\ref{fig:lumsAngle}). This contrasts with the
$U/U$ and $U/D$ cases  which show a clear, sustained collimation, radiating more strongly 
within a conical section orthogonal to the orbital plane.
The Poynting fluxes for all the cases, projected onto a sphere encompassing the binary,
are displayed in Fig.~\ref{fig:sqphi2Spheres} at $t \approx 1.5\rm{ms}$ before the
merger. The observed structures are much more involved than the previously studied
aligned cases.

Finally, one may wonder why the $U/D$ and $U/U$ cases considered in the previous paper
lack the periodicity shown in the cases here, since their luminosities
were previously explained in terms of stretching and reconnection.
However, it is important to note that the $U/D$ and $U/U$ cases preserve
an approximate azimuthal symmetry as the stars orbit. In other
words, the reconnection occurs continuously. In contrast,
the orbits of the cases here are marked by different relative
dipole orientations at different times in the orbits. The $U/D$ and $U/U$
cases essentially look the same for any point in the orbit.

%%%%%%%%%%%%%%%%%%%%%%%%%%%%%%%%%%%%%%%%%%%%%%%%%%%%%%%%%%%
%%%%%%%%%%%%%%%%%%%%%%%%%%%%%%%%%%%%%%%%%%%%%%%%%%%%%%%%%%%
\section{Discussion}
\label{section:discussion}
%%%%%%%%%%%%%%%%%%%%%%%%%%%%%%%%%%%%%%%%%%%%%%%%%%%%%%%%%%%
%%%%%%%%%%%%%%%%%%%%%%%%%%%%%%%%%%%%%%%%%%%%%%%%%%%%%%%%%%%

We study the interaction of magnetospheres in
binary systems in two different magnetization regimes.
The first corresponds to a strongly magnetized primary 
with a weakly magnetized companion at large
separations. 
We extend the analytic estimates for binaries with a single 
magnetized star, obtained with the unipolar induction model~\cite{2001MNRAS.322..695H,Piro:2012rq,Lai:2012qe},
to cases where both stars are magnetized. We
confirm the validity of this extension by studying a magnetized
star boosted with respect to an external magnetic field. This scenario
is a reasonable approximation to the dynamics of these binaries at large separations.

In the second regime, we explore magnetosphere
interactions in the final orbits prior to merger. 
We employ numerical simulations, solving the
general relativistic resistive magnetohydrodynamics equations to capture the complex 
dynamics of the binary and its magnetosphere. By comparing our results
with previous work~\cite{2013PhRvL.111f1105P,2013PhRvD..88d3011P}, we 
show that magnetospheric interaction extracts kinetic energy
from the system and powers a strong Poynting flux for different orientations of the
stellar magnetic dipoles.
Indeed, the electromagnetic luminosity for all configurations is sizable, giving
strong hope for the observation of electromagnetic counterparts to gravitational
wave events sourced during the late orbiting stages.
In particular, our results show that during these stages, the luminosities 
present an approximately cyclic temporal oscillation tied to the magnetic field and orbital
dynamics.
The luminosities for these misaligned configurations
(and possibly for generic ones as these represent rather
contrasting configurations) are roughly within the range defined by the previously
studied aligned $U/u$ and $U/D$ cases, as shown in Fig.~\ref{fig:lums}.

As discussed in detail in~\cite{2013PhRvD..88d3011P}, the energy
radiated during the coalescence of these binary systems can give rise to several 
promising emission channels. Indeed, the field configuration and dynamics obtained have features
clearly tied to models for non-thermal components in pulsars and related systems. 
As our studies indicate, the resulting Poynting flux is not strongly collimated---and thus could induce
isotropic emissions---and its complex 
time dependence is intimately tied to the field configurations in the stars. The basic
premises for non-thermal components in pulsars are present in binaries such
as studied here and can contribute to the spectra. In particular, standard estimates for synchrotron self-absorption 
indicate synchrotron radiation could be observed from radio to  gamma rays~\cite{1979rpa..book.....R} and bear
imprints of the oscillations observed in the luminosities. 
Further, these BNSs may power an expanding electron-positron wind with a thermal spectra that could be
observable in X-ray by ISS-Lobster at distances $10^{0-2}(B/10^{11}G)$ Mpc,
depending on the particular configuration of the magnetic moments.

\begin{figure}
	\includegraphics[width=0.8\columnwidth]{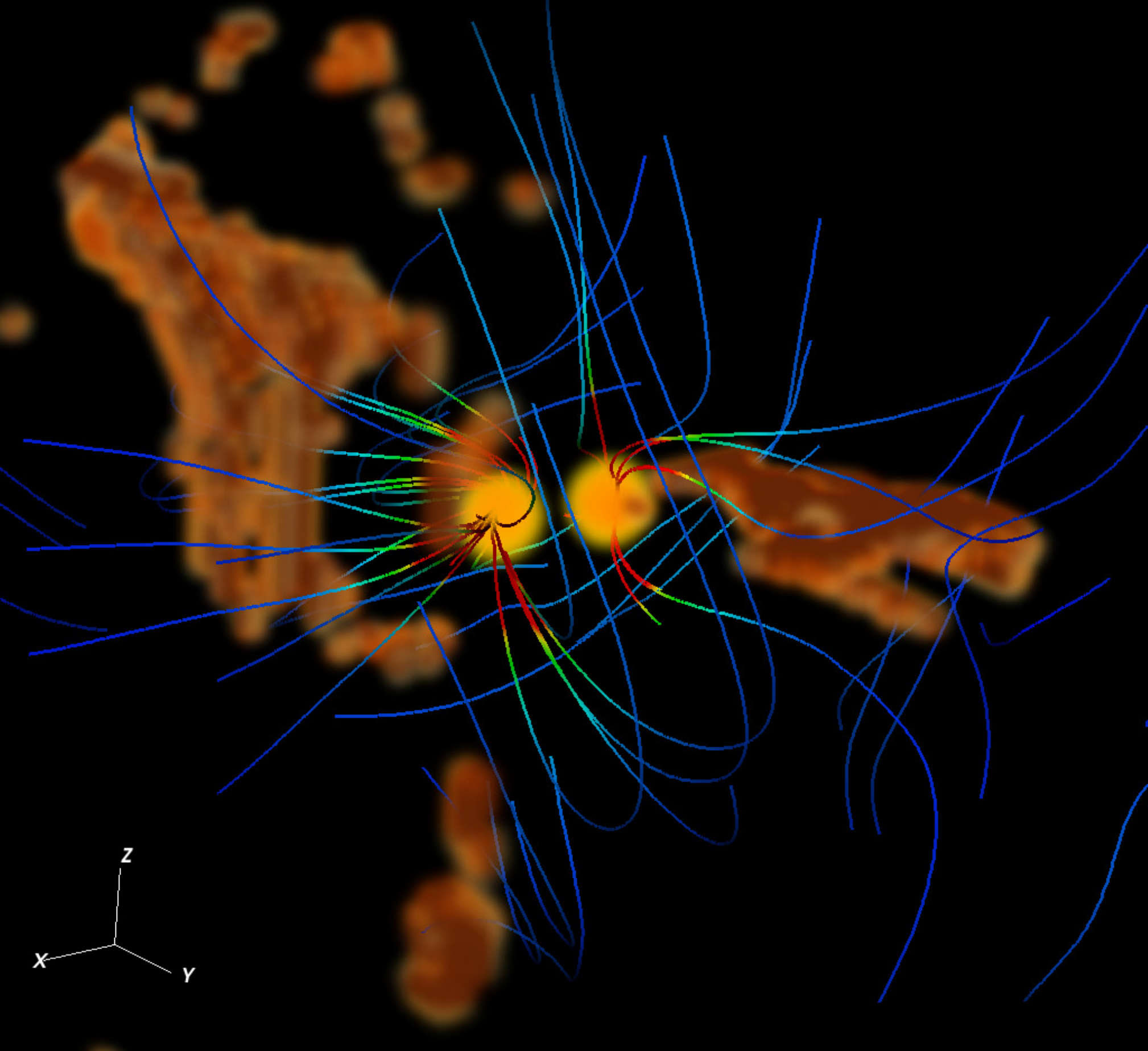}
	\\
	\includegraphics[width=0.8\columnwidth]{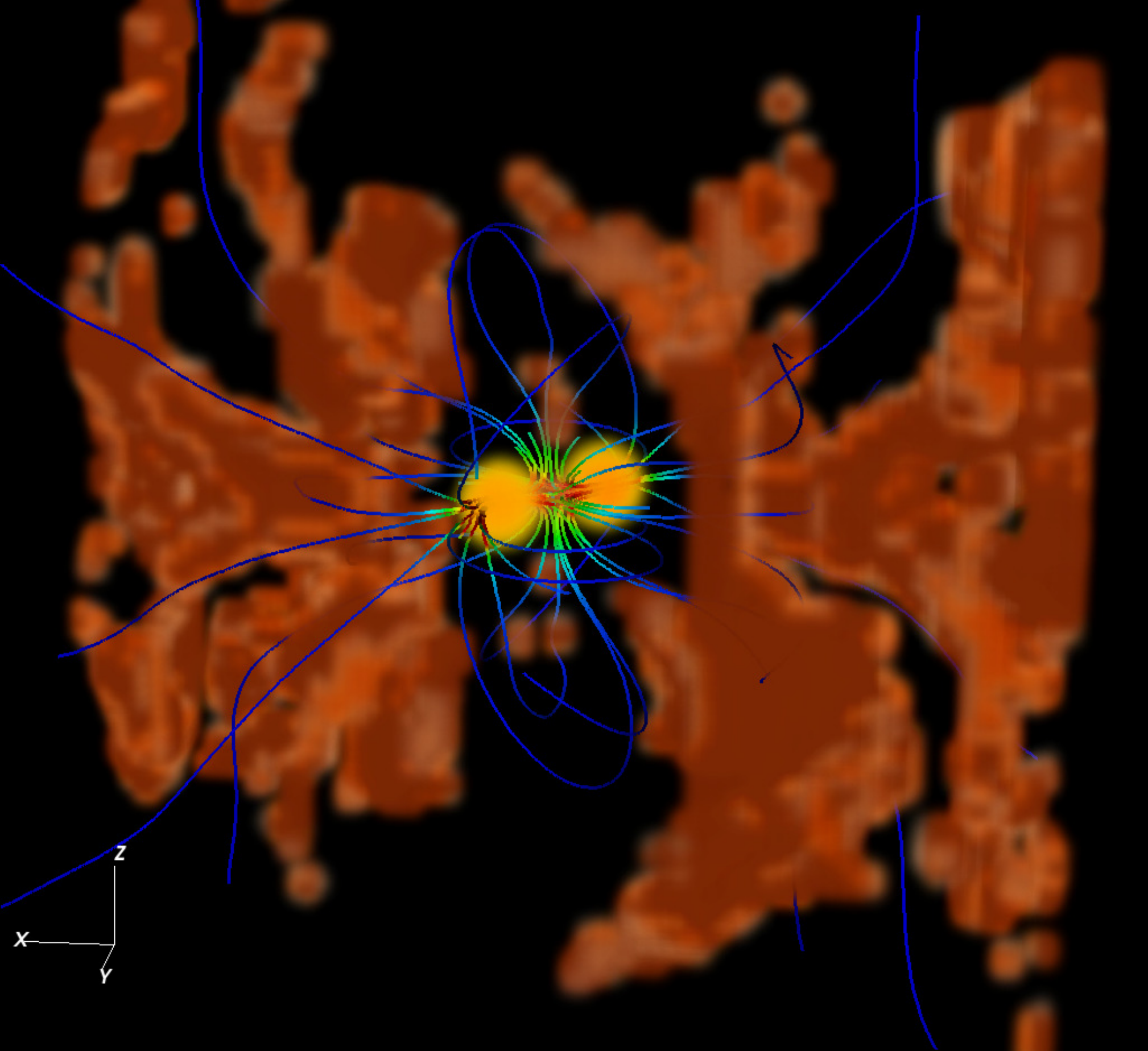}
	\caption{Current sheets in the $P/U$ (upper panel) and \textit{P/-P} (lower panel) cases,
        roughly $1.5 {\rm ms}$ before merger.
        The current sheets associated with each star appear on the plane perpendicular
        to the magnetic dipole crossing roughly through the star itself.
        Consequently, the current sheets in the
        $P/U$ case appear both along and orthogonal to the orbital plane, while 
        in the \textit{P/-P} case current sheets occur only in planes perpendicular to the orbital plane.}
	\label{fig:currentSheets}
\end{figure}

\begin{figure}
	\includegraphics[width=0.9\columnwidth]{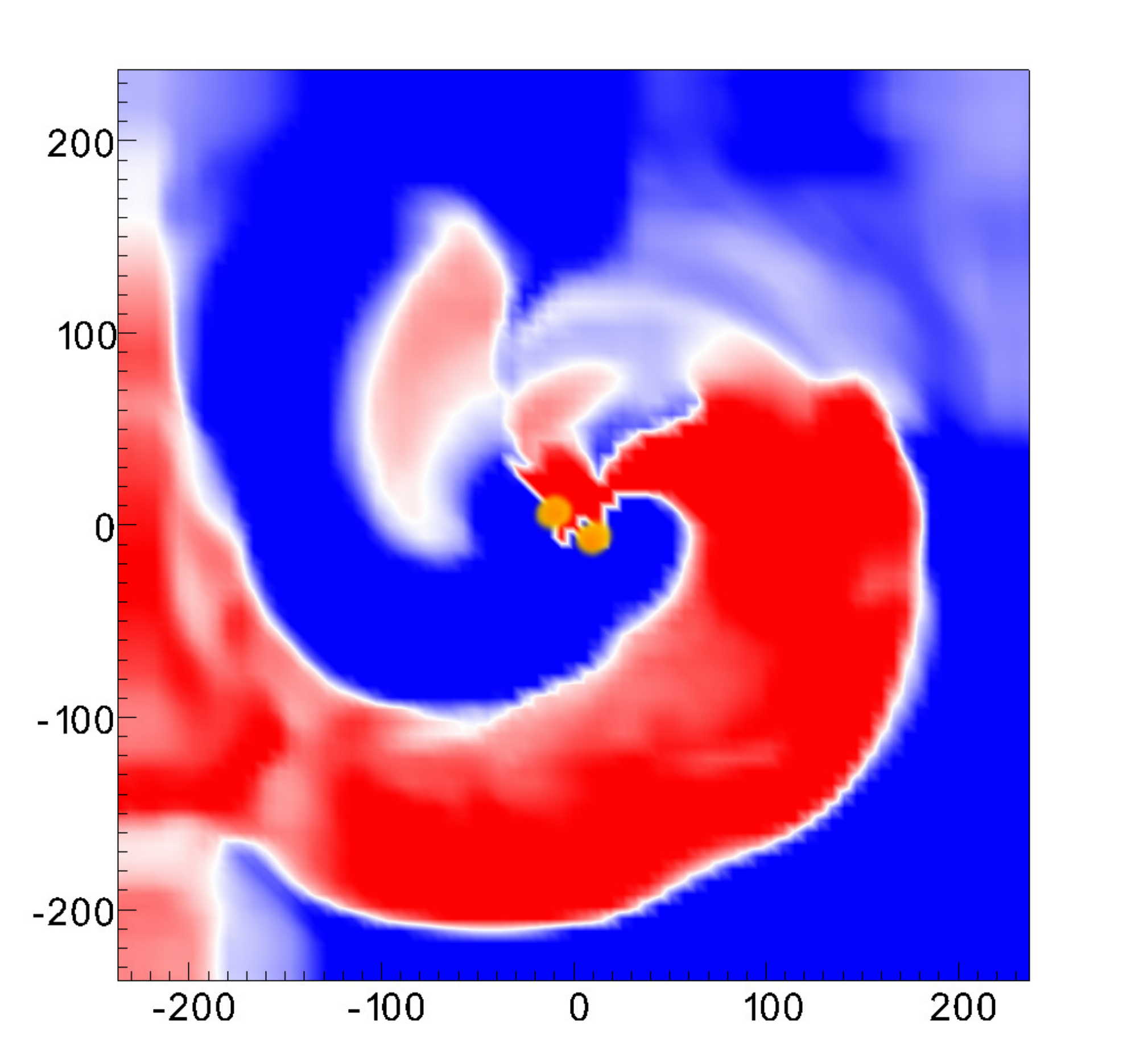}

	\caption{Polarity of the  $B_z$ component (red positive, blue negative) for the $P/U$ case,
        displayed slightly below the orbital plane ($z=-7.5 {\rm km}$), roughly $3 {\rm ms}$
        before merger. The ``striped'' pattern displays a spiral structure that
	persists during the evolution and rotates with the binary, alternating the polarity
	at any particular point approximately once per orbit.
	Similar patterns and dynamics are present for the corresponding components
        of the magnetic field in the other two orthogonal
	coordinate planes (i.e. \textit{x-z} and \textit{y-z}).
	}
	\label{fig:UP-polarity}
\end{figure}

The structure of the magnetic field lines for the misaligned
cases --except the $P/u$ case-- clearly cycles through repeated configurations
involving both stars (reconnections-stretching-deflection),
indicating that a simple model of ``electric circuits,''
with currents flowing along the magnetic field lines, might still
be useful for these systems. However, it should be stressed
that these are transitory circuits because
they are created and destroyed twice per orbit.

Indeed, the maxima in electromagnetic luminosity correlate with the reconnection events that produce a very
tight connection via magnetic flux between the stars. For the $P/P$ case 
in particular, the luminosity peaks occur when the dipoles roughly align (i.e. fall along a line)
and such an aligned configuration naturally produces a very compact, tight field configuration.
As evident in Fig.~\ref{fig:lums}, the maxima in the $P/P$ case are more luminous for those brief moments
than any of the other cases.

As in the case of isolated pulsars and aligned dipole configurations in binary systems,
current sheets\footnote{Current sheets have been linked to
different emission channels for energetic electromagnetic signals  in which synchrotron or inverse Compton
scattering may produce gamma-rays~\cite{2011SSRv..160...45U}.} also arise in the misaligned cases.
However, the character of these current sheets is quite different, being much more dynamical
and temporal, with an appearance that is recurrent with the orbital period
and  orbiting with a lag of roughly a quarter orbit with respect to the stars.
Examination of the current sheets, such as shown in Fig.~\ref{fig:currentSheets},
suggests that in the neighborhood of each star, one finds current sheets forming in
the plane orthogonal to its dipole field and roughly bisecting the star itself. For the $P/U$ case shown in the top panel,
current sheets about the star on the left form in a vertical plane whereas the star
on the right forms current sheets within the orbital plane. In the \textit{P/-P} case
(bottom panel),  both dipoles lie in the orbital plane and current sheets form in two
vertical planes that rotate with the stars.
The results of the aligned cases of Ref.~\cite{2013PhRvL.111f1105P} in which the dipoles are aligned
vertically are consistent with this suggestion; the current sheets in those cases generally arose
in the orbital plane which is the plane perpendicular to each dipole
(see Fig.~1 of Ref.~\cite{2013PhRvL.111f1105P}).

Finally, it is interesting to mention that certain features common
to both aligned and oblique rotators~\cite{Spitkovsky:2006np} clearly arise in the $P/U$ case. 
We observe an alternating polarity in the induced magnetic field configurations, see Fig.~\ref{fig:UP-polarity}.
As discussed in~\cite{2011ApJ...741...39S},
this behavior may cause strong particle acceleration and generate intense radiation via synchrotron. 
The phenomenology discussed here, together with results presented in~\cite{2013PhRvL.111f1105P,2013PhRvD..88d3011P},
explicitly argue that magnetospheric interaction in binary neutron star systems can
produce a strong electromagnetic output prior
to merger, regardless of the magnetic moments configurations.
Identifying pre-merger counterparts is important for concurrent detection because the GW frequency of the binary will increase at merger out of the sensitivity band of advanced GW detectors.

%%%%%%%%%%%%%%%%%%%%%%%%%%%%%%%%%%%%%%%%%%%%%%%%%%%%%%%%%%%%%%%%%%%%
%
%   A C K N O W L E D G M E N T S
%
%%%%%%%%%%%%%%%%%%%%%%%%%%%%%%%%%%%%%%%%%%%%%%%%%%%%%%%%%%%%%%%%%%%%
\vspace{0.5cm}

\noindent{\bf{\em Acknowledgments:}}
It is a pleasure to thank A. Broderick and C. Thompson, and our long 
time collaborators M. Anderson, E. Hirschmann, D. Neilsen and P. Motl
for useful discussions about this subject. This work was supported by the NSF under
grants PHY-0969827~(LIU) PHY-1308621~(LIU), 
NASA's ATP program through grant NNX13AH01G,
NSERC through a Discovery Grant (to LL) and CIFAR (to LL). 
C.P acknowledges support by the Jeffrey L.~Bishop Fellowship.
Research at Perimeter Institute is supported through Industry Canada and by the
Province of Ontario through the Ministry of Research \& Innovation. 
Computations were performed on the gpc supercomputer at the SciNet HPC Consortium.
SciNet is funded by: the Canada Foundation for Innovation under the auspices of
Compute Canada; the Government of Ontario; Ontario Research Fund - Research Excellence;
and the University of Toronto.

%%%%%%%%%%%%%%%%%%%%%%%%%%%%%%%%%%%%%%%%%%%%%%%%%%%%%%%%%%%%%%%%%%%%%%%%%%%%%%%%%%
%\bibliographystyle{ieeetr}
% this provides more info and also links to papers which is convenient:
%\bibliographystyle{utphys}
\bibliography{./BNStilted}

\end{document}